\documentstyle [12pt,
epsf] {article}
\newcommand{\nwc}{\newcommand}
\nwc{\cl}  {\clubsuit}
\nwc{\di}  {\diamondsuit}
\nwc{\sps}  {\spadesuit}

\nwc{\fs}  {\footnotesize\em}
\nwc{\ns}  {\normalsize}

\nwc{\hyp} {\hyphenation}
\nwc{\be}  {\begin{equation}}
\nwc{\ee}  {\end{equation}}
\nwc{\ba}  {\begin{array}}
\nwc{\ea}  {\end{array}}
\nwc{\bdm} {\begin{displaymath}}
\nwc{\edm} {\end{displaymath}}
\nwc{\bea} {\be\ba{rcl}}
\nwc{\eea} {\ea\ee}
\nwc{\ben} {\begin{eqnarray}}
\nwc{\een} {\end{eqnarray}}
\nwc{\bda} {\bdm\ba{lcl}}
\nwc{\eda} {\ea\edm}
\nwc{\bc}  {\begin{center}}
\nwc{\ec}  {\end{center}}
\nwc{\ds}  {\displaystyle}
\nwc{\bmat}{\left(\ba}
\nwc{\emat}{\ea\right)}
\nwc{\non} {\nonumber}
\nwc{\bib} {\bibitem}
\nwc{\lra} {\longrightarrow}
\nwc{\Llra}{\Longleftrightarrow}
\nwc{\ra}  {\rightarrow}
\nwc{\Ra}  {\Rightarrow}
\nwc{\lmt} {\longmapsto}
\nwc{\prl} {\partial}
\nwc{\iy}  {\infty}
\nwc{\ol}  {\overline}
\nwc{\ul}  {\underline}
\nwc{\hm}  {\hspace{3mm}}
\nwc{\lf}  {\left}
\nwc{\ri}  {\right}
\nwc{\lm}  {\limits}
\nwc{\lb}  {\lbrack}
\nwc{\rb}  {\rbrack}
\nwc{\ov}  {\over}
\nwc{\pr}  {\prime}
\nwc{\nnn} {\nonumber \vspace{.2cm} \\ }
\nwc{\Sc}  {{\cal S}}
\nwc{\Lc}  {{\cal L}}
\nwc{\Rc}  {{\cal R}}
\nwc{\Dc}  {{\cal D}}
\nwc{\Oc}  {{\cal O}}
\nwc{\Cc}  {{\cal C}}
\nwc{\Pc}  {{\cal P}}
\nwc{\Mc}  {{\cal M}}
\nwc{\Ec}  {{\cal E}}
\nwc{\Fc}  {{\cal F}}
\nwc{\Hc}  {{\cal H}}
\nwc{\Kc}  {{\cal K}}
\nwc{\Xc}  {{\cal X}}
\nwc{\Gc}  {{\cal G}}
\nwc{\Zc}  {{\cal Z}}
\nwc{\Nc}  {{\cal N}}
\nwc{\fca} {{\cal f}}
\nwc{\xc}  {{\cal x}}
\nwc{\Ac}  {{\cal A}}
\nwc{\Bc}  {{\cal B}}
\nwc{\Uc}  {{\cal U}}
\nwc{\Vc}  {{\cal V}}
\nwc{\Th} {\Theta}
\nwc{\th} {\theta}
\nwc{\vth} {\vartheta}
\nwc{\eps}{\epsilon}
\nwc{\si} {\sigma}
\nwc{\Gm} {\Gamma}
\nwc{\gm} {\gamma}
\nwc{\bt} {\beta}
\nwc{\La} {\Lambda}
\nwc{\la} {\lambda}
\nwc{\om} {\omega}
\nwc{\Om} {\Omega}
\nwc{\dt} {\delta}
\nwc{\Si} {\Sigma}
\nwc{\Dt} {\Delta}
\nwc{\al} {\alpha}
\nwc{\vph}{\varphi}
\nwc{\zt} {\zeta}
\def\tr{\mathop{\rm tr}}
\def\Tr{\mathop{\rm Tr}}

\def\VEV#1{\left\langle #1\right\rangle}

\def\abs#1{\left| #1\right|}
\def\pr#1{#1^\prime}

\nwc{\Id}  {{\bf 1}}
\nwc{\sgn}  {{\rm sgn}}
\nwc{\diag} {{\rm diag}}
\nwc{\inv}  {{\rm inv}}
\nwc{\mod}  {{\rm mod}}
\nwc{\hal} {\frac{1}{2}}
\nwc{\tpi}  {2\pi i}

\def\slash#1{#1\!\!\!/\!\,\,}

\def\pr#1{Phys. Rev. {\bf #1}}

\def\MeV {\,{\rm  MeV}}
\def\GeV {\,{\rm  GeV}}

\def \lta {\mathrel{\vcenter
     {\hbox{$<$}\nointerlineskip\hbox{$\sim$}}}}
\def \gta {\mathrel{\vcenter
     {\hbox{$>$}\nointerlineskip\hbox{$\sim$}}}}
\newsavebox{\nnin} \sbox{\nnin}{$\hspace{1mm}\in\kern -.8em /
                   \hspace{1mm}$}

\newcommand{\sub}{\subset}
\newsavebox{\nnsub} \sbox{\nnsub}{$\hspace{1mm}\sub\kern -.9em /
            \hspace{1mm}$}

\def\KK{{\rm I\kern -.2em  K}}
\def\NN{{\rm I\kern -.16em N}}
\def\RR{{\rm I\kern -.2em  R}}
\def\ZZ{Z \kern -.43em Z}
\def\QQ{{\rm \kern .25em
             \vrule height1.4ex depth-.12ex width.06em\kern-.31em Q}}
\def\CC{{\rm \kern .25em
             \vrule height1.4ex depth-.12ex width.06em\kern-.31em C}}
\def\ZZZ{Z\kern -0.31em Z}

\nwc{\olsi}   {\ol{\si}}
\nwc{\olsinuc}{\ol{\si}_{\rm nuc}}
\nwc{\olsiqm} {\ol{\si}_{\rm qm}}
\nwc{\Phid}   {\Phi^\dagger}

\newcounter{app}
\def\app{\par
 \addtocounter{app}{1}
 \def\thesection{\Alph{app}}
 \def\ksection{\Alph{app}}}

\def\appendix#1{\app\sect{#1}}

\newcommand{\sect}[1]{ \section{#1} \setcounter{equation}{0} }

\begin{document}
\bibliographystyle{physics}

\begin{titlepage}

  \title{\vspace{1cm}Quark and Nuclear Matter\\
    in the\\
    Linear Chiral Meson Model\thanks{This work is supported in part
    by funds provided by the U.S. Department of Energy (D.O.E.)
    under cooperative research agreement \# DE-FC02-94ER40818 and by
    the Deutsche Forschungsgemeinschaft.}}

  \author{ {\sc J.~Berges\thanks{Email:
        Berges@ctp.mit.edu, current address: Institut f\"ur Theoretische
Physik, Universit\"at Heidelberg, Philosophenweg 16, 69120 Heidelberg,
Germany }}\\[3mm]
    {\em Center for Theoretical Physics}\\
    {\em Massachusetts Institute of Technology}\\
    {\em Cambridge, MA 02139, USA}\\[6mm]
    {\sc D.--U.~Jungnickel\thanks{Email:
        D.Jungnickel@thphys.uni-heidelberg.de},
      C.~Wetterich\thanks{Email: C.Wetterich@thphys.uni-heidelberg.de}}
    \\[3mm]
    {\em Institut f\"ur Theoretische Physik} \\ {\em Universit\"at
      Heidelberg} \\ {\em Philosophenweg 16} \\ {\em 69120 Heidelberg,
      Germany} }

\date{}
\maketitle

\begin{picture}(5,2.5)(-350,-500)
\end{picture}

\thispagestyle{empty}

\begin{abstract}
  We present an analytical description of the phase
  transitions from a nucleon gas to nuclear matter and from nuclear matter
  to quark matter within the same model.
  The equation of state for quark and nuclear matter is encoded in the
  effective potential of a linear sigma model.  We exploit an exact
  differential equation for its dependence upon the chemical potential
  $\mu$ associated to conserved baryon number. An approximate solution 
  for vanishing temperature is used to discuss possible phase transitions
  as the baryon density increases. For a nucleon gas and nuclear matter we
  find a substantial density
  enhancement as compared to quark models which neglect
  the confinement to baryons. The results point out
  that the latter models are not suitable to discuss the 
  phase diagram at low temperature. 
\end{abstract}

\end{titlepage}

\section{Introduction}

The equation of state for strongly interacting matter at nonzero
density is needed for the understanding of neutron stars \cite{neutron},
as well as for the interpretation of heavy ion collision
experiments \cite{heavyion}.
Any analytical description of the equation of state for baryons at nonzero
baryon density has to cope with the problem that the effective degrees of
freedom change from nucleons at low density to quarks at high density.  We
attempt in this note a unified description of both the nuclear gas--liquid
transition and the transition to quark matter. For this purpose we work
within an effective linear meson model coupled to quarks and nucleons.  It
should describe the low momentum degrees of freedom of QCD for the range of
temperatures and densities which are relevant for these phase transitions.
Our main computational tool will be a new exact functional differential
equation for the dependence of the effective action on the baryon chemical
potential and an approximate solution to it. We will see many similarities
but also important differences as compared to a mean field theory
treatment.

Our main interest are the chiral aspects of
the equation of state for quark and nuclear matter at
nonzero baryon density and the order of the involved phase transitions.
In this note we concentrate on the simplest possibility where only the
color-singlet chiral condensate plays a role in the transitions.
Our investigation should constitute a useful point
of comparison for other models with more complicated condensates.
In fact, in addition to the nuclear and quark matter
phases a number of interesting
possibilities like the formation of meson condensates or strange quark
matter~\cite{Bod71-1,Wit84-1,FJ84-1} have been proposed. Also an
extensive discussion has focused around the symmetry of the high density
state, where the spontaneous breaking of color is
associated with the phenomenon of color superconductivity
\cite{BL84-1,ARW98-1,RSSV98-1,CSC}.  Other ideas concern
the spontaneous breaking of the color symmetry in
the vacuum \cite{W3F,BW2F}. In this note we adopt the working
assumption that possible additional condensates have only little influence
on the transitions associated with the order parameter of chiral symmetry
breaking\footnote{The results of \cite{BR98-1} indicate that
  condensates of quark Cooper pairs do not influence the behavior of the
  chiral condensate to a good approximation. The phenomenon of
  high density
  color superconductivity has only minor influence on the equation
  of state for quark matter \cite{CSC}. On the other hand, for
  a substantial color octet condensate in the vacuum \cite{W3F,BW2F}
  our assumption would not hold.}.  We
also concentrate mainly on the case of two quark flavors and neglect
isospin violation\footnote{Isospin violation and electromagnetism are
  important for nuclear matter in neutron stars. Our result for the
  equation of state is therefore not quantitatively realistic in all
  respects.  Isospin violation can be incorporated in our formalism without
  conceptual difficulties. We have already included electromagnetism
  phenomenologically for the quantitative description of nuclei.}.

Within this setting and using a rather crude approximation for the transition
from quark to nucleon degrees of freedom, we
find that for low temperature both the nuclear gas--liquid and the
hadron--quark transitions are of
first order, in accordance with indications from earlier
investigations~(cf.~\cite{Kle92-1,BR98-1,HJSSV98-1} and \cite{CSC}
and references therein). 
The phase transition from nuclear to quark matter tends to be
much stronger (larger surface tension) than the gas--liquid nuclear
transition. The first order character of these phase transitions would have
important implications. In particular, one may combine this with
information about the high temperature phase transition for vanishing
baryon density: One expects~\cite{BCCGP,BR98-1,HJSSV98-1} an endpoint of the
first order critical line between quark and nuclear matter if the zero
density, high temperature transition is a crossover (as for two flavor QCD
with non--vanishing quark masses). Such an endpoint corresponds to a second
order transition where a large correlation length may lead to
distinctive signatures in relativistic heavy ion collisions~\cite{SRS98-1}.
If the zero density transition turns out to be of first order in three
flavor QCD, such an endpoint does not necessarily occur.  (Endpoints are
not excluded in this case, however, since the two first order regions could
be disconnected.)  The first order line for the gas--liquid nuclear
transition exhibits a critical endpoint for a temperature of about
$10\MeV$.  Signatures and critical properties of this point have been
studied through measurements of the yields of nuclear fragments in low
energy heavy ion collisions \cite{CK86-1,EMU98-1}. 

We emphasize that our treatment of quarks and baryons is still very crude
and a different picture of the high density transition, e.g.\ due to the
inclusion of other condensates, may well appear for two flavor QCD.
A general outcome of our analysis concerns the crucial importance of 
confinement for any understanding
of the phase transitions at high density and low temperature. In fact,
the contribution of a free gas of baryons to the dependence of the free
energy on the chemical potential is enhanced by a factor 27 (!) as compared 
to the contribution from a free gas of quarks. Therefore, the binding of 
quarks into baryons at low density and temperature
plays a crucial quantitative role which cannot be neglected by
any satisfactory treatment of the high density transition. In fact,
a first order transition between nuclear matter and quark matter would
presumably connect an approximately free nucleon gas at low density
to a quark gas at high density. Such a transition involves then a
description in terms of baryons in the low density phase and cannot
be understood within quark descriptions, which do not reflect the
large baryonic enhancement factor. This may explain the phenomenologically
unacceptable low critical densities often found in such quark
descriptions.

In quantum field theory the effects of a non--vanishing baryon density in
thermal equilibrium or the vacuum are described by adding to the classical
action a term proportional to the chemical potential $\mu$,
\begin{equation}
  \label{AAA001}
  \Delta_\mu S=3i\mu\sum_j b_j
  \int_0^{1/T}d x_0
  \int d^3 \vec{x}\ \ol{\psi}_j\gm^0\psi_j\equiv
  -3\frac{\mu}{T}B\; .
\end{equation}
The index $j$ labels all fermionic degrees of freedom which carry a
non--vanishing baryon number $b_j$ and a summation over spinor indices is
assumed implicitly.  For a description of the fermionic degrees of freedom
in terms of quarks the sum is over $N_{\rm c}$ colors and $N_{\rm F}$
flavors, with $b_j=1/3$.
We neglect the heavy quarks and concentrate on a two-flavor
approximation where also the strange quark is omitted.
For the nucleon degrees of freedom we include
protons and neutrons with $b_j=1$.  For our conventions,
$\mu$ corresponds to the chemical potential of quark number density.  The
baryon number density $n$ can be obtained from the $\mu$--dependence of the
Euclidean effective action $\Gamma$, evaluated at its minimum for fixed
temperature $T$ and volume\footnote{More precisely, $B$ counts the number
  of baryons minus antibaryons. For $T\ra0$ the factor $T/V$ is simply the
  inverse volume of four--dimensional Euclidean space.} $V$
\begin{equation}
  \label{AAA002}
  n\equiv\frac{\VEV{B}}{V}=
  -\frac{1}{3}\frac{\prl}{\prl\mu}
  \left.\frac{\Gamma_{\rm min}T}{V}\right|_{\rm T,V}\; .
\end{equation}
We note that the Helmholtz free energy is $F=\Gm_{\rm min}T+3\mu n V$.  Our
aim is a computation of the difference of $\Gamma_{\rm min}$ between
non--vanishing and vanishing $\mu$. For $T=0$ this is dominated by
fermionic fluctuations with (spatial) momenta $\vec{q}^{\,2}\leq\mu^2$.
For not too large $\mu$ (say $\mu\lta\,600\, \MeV$) we can therefore work
with an effective model for the low momentum degrees of freedom of QCD.
This argument generalizes to moderate temperatures, say $T\lta200\MeV$.
In the bosonic sector we will work with a linear meson model whereas
for the fermions we keep the multiplet with lowest mass as discussed
above. Our description takes into account the lightest scalar and
pseudoscalar mesons as well as the lowest multiplet of vector mesons.

The minimum of the effective action corresponds to the minimum of the
effective meson potential $U=\Gamma \, T/V$ for constant
scalar meson fields.  In consequence, $U$
is a function of a complex $N_{\rm F}\times N_{\rm F}$ scalar field matrix
$\Phi$, which describes the nonets of scalar and pseudoscalar mesons, and a
similar matrix for the vector mesons.  For a discussion of the chiral phase
transition it will be sufficient to know the dependence of $U$ on space and
time independent fields which can acquire a vacuum expectation value
consistent with $SU(N_{\rm F})$ symmetry. These
are the real diagonal elements of $\Phi$ which we denote by $\si$,
and similar diagonal elements $\om$ for the zero component of the vector
mesons\footnote{The zero component of the vector fields $\om_\mu$ 
can aquire a nonvanishing expectation value since at nonzero chemical 
potential Lorentz invariance is broken.}.  
In the limit of vanishing current quark masses the minimum
of $U$ at sufficiently high temperature or high density 
should occur at $\si=0$ in this model.
For low $T$ and $\mu$ spontaneous chiral symmetry breaking is triggered by
a non--vanishing expectation value $\olsi(\mu,T)$, corresponding to the
location of the minimum of $U(\si,\om;\mu,T)$. (We adopt the convention
through this work that bars indicate locations of potential minima.) The
explicit breaking of chiral symmetry through non--vanishing current quark
masses is described by a linear source term contained in $U$ which induces
nonzero $\olsi$ even in the 
phase without spontaneous symmetry breaking\footnote{For general quark or
nucleon masses the diagonal elements $\si$ of the 
scalar meson matrix, and similarly for $\om$, can differ from each other.
We suppress this dependence in the notation since we will only 
perform calculations for the case where all diagonal elements are 
the same.}~\cite{PW84-1,JW95-1,BJW98-1}.  
The baryon density $n$, energy
density $\eps$ and pressure $p$ follow from $U(\mu,T)\equiv
U(\olsi(\mu,T),\ol{\om}(\mu,T);\mu,T)=\eps-T s-3\mu n$ as
\begin{eqnarray}
  \label{AAA003}
  \ds{n} &=& \ds{
    -\frac{1}{3}\frac{\prl}{\prl\mu}
    U(\mu,T)\; ;\;\;\;
    p=-U(\mu,T)
    }\nnn
  \ds{\eps} &\equiv& \ds{
    \frac{E}{V}=
    U(\mu,T)+3\mu n-
    T\frac{\prl U}{\prl T}(\mu,T)
    }\; .
\end{eqnarray}
Here we have normalized $U(0,0)=0$ corresponding to vanishing
pressure in the vacuum.

For fluctuations in the momentum range $q_{\rm
  H}^2<\vec{q}^{\,2}<(600\MeV)^2$ we work within the linear quark meson
model~\cite{JW95-1,BJW98-1,GM84-1,JW98-1,JW96-1,JW98-2} in an approximation
which does not describe the effects of confinement.  For low momenta,
i.e.~$\vec{q}^{\,2}<q_{\rm H}^2$, this description therefore becomes
inappropriate. Three quarks are bound into color singlet nucleons. In this
momentum range we describe the fermionic degrees of freedom by baryons,
while keeping the description of the bosons in terms of the scalar field
$\Phi$ and corresponding vector meson fields. The use of the same bosonic
fields for the whole momentum range will turn out to be an important
advantage since it facilitates the computation of the free energy in
different ranges of $\mu$, corresponding in turn to different baryon
densities and a different picture for the relevant fermionic degrees of
freedom. For nuclear matter a typical value of the ``transition momentum''
is $q_{\rm H}\gta260\MeV$.  We find that the quark--hadron phase transition
is substantially influenced by the change from quark to baryon fields at
$q_{\rm H}$. This implies that a reliable quantitative understanding
of this transition requires also a quantitative treatment of the 
change of effective fermionic degrees of freedom.
 
Furthermore, our computation reveals that 
the transition from a nucleon gas to nuclear matter can be
described realistically in terms of nucleon and meson
degrees of freedom only if one accepts a relatively complicated
form of the vacuum effective potential for the color-singlet
chiral order parameter $\sigma$. In particular, one needs large
higher order couplings which do not seem very natural.
It is conceivable that this situation changes for a more complex 
vacuum with additional condensates.

\section{Chemical potential flow equation}

We employ a new method for the computation of the $\mu$--dependent part of
the effective action that relies on an exact functional differential
equation for $\Gamma$. This equation expresses the $\mu$--derivative of
$\Gamma$ in terms of the exact field dependent fermion propagator. We start
from the generating functional of the connected Green functions
\begin{equation}
  \label{AAA003a}
  W[\jmath]=\ln \int D\chi
  \exp\left\{-S[\chi]-\Delta_\mu S[\chi]+\int\jmath\chi\right\}
\end{equation}
where $\chi$ stands collectively for bosonic and fermionic fields with
associated sources $\jmath$ and $S$ is the action for $\mu=0$. For our
purpose it is convenient to subtract from the effective action (defined by
a Legendre transform) the $\mu$--dependent fermion bilinear~(\ref{AAA001}):
\begin{equation}
  \label{AAA003b}
  \Gamma[\varphi]=-W[\jmath]+\int\jmath\varphi-\Delta_\mu S[\varphi],\quad
  \varphi=\frac{\delta W}{\dt\jmath} \; .
\end{equation}
The $\mu$--dependence of $\Gamma$ arises only through $\Delta_\mu S$ and
can be expressed by a trace over the connected two--point function. Using
the manipulations of generating functions outlined in~\cite{Wet90-1} in a
context with fermions~\cite{Wet93-3,BW93-1,BTW} one obtains the exact
nonperturbative functional differential equation\footnote{We mention that
  in the presence of a local gauge symmetry this equation is manifestly
  gauge invariant.}
\begin{equation}
  \label{AAA004}
  \frac{\prl}{\prl\mu}\Gamma=
  -\Tr\left\{\frac{\prl R_\mu}{\prl\mu}
  \left(\Gamma^{(2)}+R_\mu\right)^{-1}\right\}
\end{equation}
where
\begin{equation}
  \label{AAA005}
  R_{\mu,j j^\prime}(q,q^\prime)=
  3i\mu b_j\gm^0(2\pi)^4\dt(q-q^\prime)
  \dt_{jj\prime}\; .
\end{equation}
We remind that $\Gamma$ is a functional of the meson and fermion fields,
and the $\mu$--derivative on the left hand side of~(\ref{AAA004}) is taken
for fixed fields. The exact inverse propagator $\Gamma^{(2)}$ is the second
functional derivative with respect to the fields. It is a matrix in the
space of internal indices and momenta and involves fermions and bosons.
Since $\Delta_\mu S$ only affects fermions, the trace is over fermionic
indices only and contains a momentum integration. For a configuration with
constant bosonic fields and vanishing fermion fields $\Gamma^{(2)}$ does
not mix bosons and fermions and is diagonal in momentum space. We therefore
only need the inverse fermion propagator
\begin{equation}
  \label{AAA006a}
  \Gm_{j j^\prime}^{(2)}(q,q^\prime)=
  H_{j j^\prime}(q)(2\pi)^4
  \dt(q-q^\prime)
\end{equation}
in order to obtain an exact equation for the $\mu$--dependence of the
effective potential
\begin{equation}
  \label{AAA006b}
  \frac{\prl U}{\prl\mu}=
  -\sum_{j}3b_j
  \int\frac{d^4 q}{(2\pi)^4}
  \tr i\gm^0
  \left[ H(q)+3i b\mu\gm^0\right]^{-1}_{j j}\; .
\end{equation}
Here $\tr$ denotes a Lorentz trace and $H_{j j^\prime}$, $b_{j
  j^\prime}=b_j\dt_{j j^\prime}$ are matrices in the space of fermion
species. This exact relation expresses the baryon density for arbitrary quark
mass term\footnote{The relation (\ref{AAA006b}) holds for
arbitrary $\sigma$ which corresponds to arbitrary quark
mass $m_{q}$ through $\partial U/\partial\sigma\sim m_{q}$.}
(cf. eq. (\ref{AAA031})) in terms of the exact fermionic
propagators in presence of nonvanishing meson fields. We will
see that the momentum integral is both ultraviolet and infrared
finite such that eq. (\ref{AAA006b}) is well defined.

The exact fermion propagators are not known, and we have to proceed
to approximations. The advantage of our underlying exact expression
remains, however, that is easy to study which are the effects of
qualitative and quantitative changes in the approximations for the fermionic  
propagators. In particular, we will learn how the transition
from quark to nucleon degrees of freedom strongly affects the
form of the effective potential -- a discussion that would not be
possible within a mean field approximation for a given effective
model either of quarks or of baryons alone. In the present work
we will use a rather simple approximation both for the quark and
baryon propagators. For
arbitrary $\si$ and $\om$ we approximate
\begin{equation}
  \label{AAA006}
  H_{j j^\prime}(q)=\left[
  q_\nu\gm^\nu+m_j(\si;\mu,T)\gm^5+
  i b_j\Om
  \gm^0\right]
  \dt_{jj^\prime}
\end{equation}
with
\begin{equation}
  \label{eq:AAA001}
  \Om= g_{\om}(\si,T)\, \om \, .
\end{equation}
We use our
ansatz for the fermion propagator only to compute the $\mu$--dependent
contributions to the effective potential, i.e.~we consider here the
difference $U(\si,\om;\mu,T)-U(\si,\om;0,T)$.  In fact, the computation of
the contributions due to a non--vanishing chemical potential 
allows one to use quite crude approximations in many situations. 
This is based on
the observation that strongly interacting fermions are often successfully
described as freely propagating quasi--particles. In our case they acquire
an effective ``constituent'' mass $m \gamma_5$ through a strong Yukawa
coupling to mesonic vacuum expectation values. (The matrix $\gamma_5$
appears in the mass term as a consequence of our Euclidean
conventions~\cite{Wet93-3}.)  Similarly the constant field
$\om$ denotes the analytic continuation of the zero
component of the Euclidean $\om$--vector--meson field
with coupling $g_{\om}$ to the fermions.
The piece $\sim\Omega$ is the remnant of the vector coupling
$\sim\omega_\mu \gamma^\mu$ in a situation where the zero component
of $\om_\mu$
can acquire an expectation value due to the breaking of Lorentz 
invariance\footnote{For $\mu=0$ one therefore has $\omega=0$.} by the
nonvanishing chemical potential (\ref{AAA001}). We will later determine
the values of $\omega$ self-consistently. An important simplification
in our ansatz (\ref{AAA006}) is the neglection of a possible wave
function renormalization $Z_\psi(q,\sigma,\omega)$ which could multiply
the kinetic term $q_\nu\gamma^\nu$. In a more realistic setting this will
certainly play a role near the transition between quarks and baryons.

We emphasize that the computation of $\partial U/\partial\mu$ according
to eq. (\ref{AAA006b}) does not need any information
about the masses and effective self-interactions of the mesons. They
determine, however, the effective potential at zero baryon density
$U(\sigma,\omega;0,T)$ and therefore influence the possible phase
transitions. In fact, the
meson self--interactions
may turn out to be quite complicated.
We do not attempt here to compute the meson masses and self-interactions
by a mean field approximation, since
earlier renormalization group investigations have shown \cite{BJW98-1} that
this is probably much too crude.  The advantage of our method is that
the lack of knowledge about the meson interactions can be
separated from computation of the $\mu$-dependence of the potential.
We will simply parametrize $U(\sigma,\omega;0,T)$ in accordance with
the symmetries and (indirect) observational knowledge.

At this point it may be useful to summarize the approximations
that affect the computation of the $\mu$-dependence of the effective
potential if we use the ansatz (\ref{AAA006}) in the exact relation
(\ref{AAA006b}). Perhaps most importantly we omit
the dependence of the fermion wave
function renormalization on momentum, $\si$, $\omega$, $\mu$ and $T$.
We also neglect
a possible difference in normalization of the quark kinetic term and the
baryon number current. Similarly, we have
not considered a possible momentum dependence
of the mass term as well as the momentum dependence of the contribution
$\sim\gm^0$.  Finally, we assume that $m_j$ can be taken as independent
of $\omega$, and $g_{\om}$ as not dependent on $\omega$ or $\mu$.
In this approximation the term $\sim\Om$ can be combined
with $R_\mu$ such that $\mu$ is replaced in the propagator
$(\Gamma^{(2)}+R_\mu)^{-1}$ by an effective chemical potential
\begin{equation}\label{AA}
  \mu_{\rm eff}=\mu+
    \frac{1}{3}\Om(\om,\si;T)=\mu +\frac{1}{3}g_\om \om\; .
\end{equation}

With the approximation~(\ref{AAA006}) the evolution equation for the
$\mu$--dependence of the effective meson potential takes a very simple
form
\begin{equation}
  \label{AAA007}
  \frac{\prl U}{\prl\mu}=-\sum_{\rm j}3b_j
  \int\frac{d^4 q}{(2\pi)^4}\tr\left\{
  i\gm^0\left(\slash{q}+m_j\gm^5+
  3ib_j\mu_{\rm eff}\gm^0\right)^{-1}\right\}\; .
\end{equation}
The remaining trace over spinor indices is easily performed
\begin{eqnarray}
  \label{AAA008}
  \ds{\frac{\prl U}{\prl\mu}} &=& \ds{
    -2\sum_{\rm j}
    \int\frac{d^3\vec{q}}{(2\pi)^3}K_j
    }\; , \\[2mm]
  \ds{K_j} &=& \ds{
    6i b_j\int\frac{d q^0}{2\pi}
    \frac{(q^0+3i b_j\mu_{\rm eff})}
    {\left[(q^0+3ib_j\mu_{\rm eff})^2+\vec{q}^{\,2}+m_j^2\right]}
    }\; .
\end{eqnarray}
For non--vanishing temperature the $q^0$--integration is replaced by a
sum over Matsubara frequencies
\begin{equation}
  \label{AAA009}
  \int\frac{d q^0}{2\pi}\longrightarrow
  T\sum_{n\in\ZZZ}
\end{equation}
with $q^0=2\pi(n+1/2)T$ and, correspondingly,
$\dt(q-q^\prime)\ra\dt(\vec{q}-\vec{q}^{\,\prime})\dt_{n n^\prime}/(2\pi
T)$. This results in
\begin{equation}
  \label{AAA200}
  K_j=3b_j\left(
  \Big[ e^{\frac{\sqrt{\vec{q}^{\,2}+m_j^2}-
      3b_j\mu_{\rm eff}}{T}}+1\Big]^{-1}-
  \Big[ e^{\frac{\sqrt{\vec{q}^{\,2}+m_j^2}+
      3b_j\mu_{\rm eff}}{T}}+1\Big]^{-1}
  \right)
\end{equation}
where the two terms are proportional to the fermion and anti--fermion
contributions to $n$. We see explicitly that the momentum integration
is finite due to the exponential suppression for large $\vec q^2$.

We will concentrate here mainly on $T=0$ where the $q^0$--integration
yields a step function:
\begin{equation}
  \label{AAA010}
  K_j=3b_j\Theta(9b_j^2(\mu_{\rm eff})^2-
  (\vec{q}^{\,2}+m_j^2))\; .
\end{equation}
The remaining $\vec{q}$--integration is therefore cut off in the ultraviolet,
$\vec{q}\,^2<9b_j^2(\mu_{\rm eff})^2-m_j^2$, and only involves
momenta smaller than the Fermi energy $3b_j\mu_{\rm eff}^{\rm(j)}$. As it
should be, it is dominated by modes with energy $\sqrt{\vec{q}\,^2+m_j^2}$
near the Fermi surface.

\section{Quark and nucleon degrees of freedom}

We will assume that $\partial U/\partial\mu$ can be expressed as a simple
sum of the contribution from quarks with momenta $\vec{q}\,^2>q^2_{\rm H}$
and that of baryons with momenta $\vec{q}\,^2<q_{\rm H}^2$. This is the
simplest approximation which catches the effective transition from quarks to
baryons as effective degrees of freedom in the relevant momentum range.
It will be sufficient to demonstrate the most important effects
of confinement on the $\mu$-dependence of the effective potential,
namely that the contribution of a gas of nucleons is greatly enhanced
as compared to a corresponding contribution of quarks. In a more
realistic scenario the transition between quark and nucleon degrees
of freedom will be less abrupt. Within our approximations part of
the uncertainty related to this effective transition can be studied
by allowing that $q_H$ depends on $\sigma$ since typically
the relevant values for $\sigma$ depend on the baryon density,
being higher for a nucleon gas than for a quark gas.

We first
consider the range of momenta with $\vec{q}^{\,2}\ge q_{\rm H}^2$ for which
we use an effective linear quark meson
model~\cite{JW95-1,BJW98-1,GM84-1,JW98-1,JW96-1,JW98-2}. Here the quark
mass term $\sim\gm^5$ arises through a Yukawa coupling $h$ to the
expectation value of the $\si$--field, $m_q=h_q(\si;\mu,T)\si$. 
Since the quark description
breaks down for small momenta, we restrict the integration over
$\vec{q}\,^2$ in~(\ref{AAA008}) to the range $\vec{q}\,^2>q^2_{\rm H}$. We
therefore infer for the quark contribution to the $\mu$--dependence of the
effective potential (for $\mu>0$ and $T=0$)
\begin{equation}
  \label{AAA011}
  \frac{\prl U^{\rm(Q)}}{\prl\mu}=
  -\frac{N_{\rm c} N_{\rm F}}{3\pi^2}
  \left[\left(
  \mu_{\rm eff}^{2}-h^2_q\si^2\right)^{3/2}
  -q^3_{\rm H}\right]
  \Theta\left(\mu_{\rm eff}^{2}-
  h^2_q\si^2-q^2_{\rm H}\right)\; .
\end{equation}

For the low momentum range $\vec{q}\,^2<q_{\rm H}^2$ where the fermionic
degrees of freedom are the lightest baryons rather than quarks we repeat
the steps leading from~(\ref{AAA004}) to~(\ref{AAA011}). The trace now
involves a sum over proton and neutron but no color factor.  
This yields a contribution (again for
$T=0$)
\begin{eqnarray}
  \label{AAA014a}
  \ds{\frac{\prl U^{\rm(B)}}{\prl\mu}} &=& \ds{
    -\frac{2}{\pi^2}
    \Bigg\{(9\mu_{\rm eff}^{2}-m_N^2)^{3/2}
    \Theta(9\mu_{\rm eff}^{2}-m_N^2)
    \Theta(m_N^2+q_{\rm H}^2-9\mu_{\rm eff}^{2})
    }\nnn
  &+& \ds{
    q^3_{\rm H}\Theta(9\mu_{\rm eff}^{2}-m_N^2-q_{\rm H}^2)\Bigg\}
    } \, .
\end{eqnarray}
We parametrize the nucleon mass as $m_N(\sigma) = 3 h_N(\sigma) \sigma$
and note that for $h_N \simeq h_q$ one has $m_N \simeq 3 m_q$, 
as appropriate for nucleons described as composites of three 
constituent quarks.
The baryon density can be directly inferred from eqs.~(\ref{AAA011}),
(\ref{AAA014a}) as
\begin{equation}
  \label{eq:AAA002}
  n=-\frac{1}{3}\left.\left(\frac{\prl U^{\rm(Q)}}{\prl\mu}+
  \frac{\prl U^{\rm(B)}}{\prl\mu}\right)
  \right|_{\si=\olsi,\;\om=\ol{\om}}
\end{equation}
since the partial derivatives of the effective potential with respect to
$\si$ and $\om$ vanish at the $\mu$--dependent potential minimum
$(\olsi,\ol{\om})$.

We repeat that this picture is only a crude approximation to the
binding of quarks into nucleons. A nucleon description should work well for
$h_{\rm N}^2\si^2$ near $\mu_{\rm eff}^2$, since only low momentum degrees
of freedom contribute in this range.  On the other hand, the quark
description becomes important for $h^2\si^2\ll \mu_{\rm eff}^2-q^2_{\rm
 H}$.  In a more realistic description the $\Theta$--functions
in~(\ref{AAA011}), (\ref{AAA014a}) would become smooth. The
characteristic quark--baryon transition momentum $q_{\rm H}$
typically depends on
$\si$. Indeed, a baryon description for the low momentum degrees of freedom
is necessary for $\si$ not too far from its vacuum expectation value
$\si_0$.  We will see that in this range of $\sigma$ the transition
momentum $q_H$ is around 260 MeV or higher.
On the other hand, baryons do not seem to be meaningful degrees
of freedom in a situation of chiral symmetry restoration at $\si=0$
such that $q_H$ may vanish for $\sigma=0$.

For $h_{\rm N}$ of the same order as $h$, the nucleon mass is about
three times the quark mass for a given value of $\sigma$. Therefore
eq.~(\ref{AAA014a}) results in an
important enhancement of $\prl U/\prl\mu$ in the range $\sqrt{\mu_{\rm
    eff}^2-\frac{1}{9}q_{\rm H}^2}<h_{\rm N}\abs{\si}<\mu_{\rm eff}$ as
compared to the contribution from the quarks.  This is mainly
due to the fact that
more energy levels fall below the Fermi energy $3\mu_{\rm eff}$ for the
baryons.
More precisely, a factor $3^3=27$ arises from the ratio
$[(9\mu^{2}-m_N^2)/(\mu^{(a)2}_{\rm eff}-h_q^2\sigma^2)]^{3/2}$
if the subtraction of $q^3_H$ in eq. (\ref{AAA011}) can be neglected.
An additional suppression  of the quark contribution by a factor
$b_q=1/3$ in the coupling of the chemical potential is canceled by
the color factor $N_c=3$. (Neglecting strangeness, the two
quark flavors are matched by the two species of nucleons. For a light
strange quark one would observe an additional enhancement for the baryon
contribution due to the larger number of baryons in an octet as compared
to the three flavors.) This ``nucleon enhancement'' is one of
the most important observations of the present paper. We believe that
this effect is quite robust in view of a possible more precise
modeling since only very simple properties of the fermion degrees
of freedom play a role for our argument. In our description the
large ``nucleon enhancement'' by a factor of about
$27$ is the basic mechanism which may lead to separate gas--liquid and
hadron--quark phase transitions. Despite this enhancement one observes that
$\prl U/\prl\mu$ is continuous in $\si$ and $\mu$. Furthermore,
for $h_N(\sigma)=h_q(\sigma)$ the simultaneous jump of the
renormalized fermion mass by a factor of three, together with a
similar jump of the renormalized coupling to vector mesons (due to the
factor $b_j$ in eq. (\ref{AAA006})) could also be accounted for
by a sudden drop of the fermion wave function
renormalization\footnote{Such a drop of the effective wave
function renormalization would be required for a Higgs picture
of the QCD vacuum where quarks and baryons are described
by the same field \cite{W3F}.}
form $Z_\psi=1$ for $q^2>q^2_H$ to $Z_\psi=1/3$ for
$q^2<q^2_H$. This corresponds to the continuity  in $\mu_{\rm eff}$
which does not depend on the wave function renormalization
multiplying the fermion kinetic term.

With $\frac{\prl}{\prl\mu_{\rm eff}}=\frac{\prl} {\prl \mu}$ we can easily
rewrite eqs.~(\ref{AAA011}), (\ref{AAA014a}) as flow equations for $\mu_{\rm
  eff}$.  In the approximation of $\mu$--independent Yukawa couplings
$h=h(\si)$, $h_{\rm N}=h_{\rm N}(\si)$ and $q_{\rm H}=q_{\rm H}(\si)$ these
differential equations can be integrated analytically.  We define
\begin{eqnarray}
  \label{AAA012}
  \ds{U(\si,\om;\mu,T)} &\equiv& \ds{
    U_0(\si;T)+U_\om(\si,\om;T)+
    2U_\mu(\si,\om;\mu,T)
    }
\end{eqnarray}
where $2U_\mu(\si,\om;\mu,T)$ entails the $\mu$--dependent contribution
from the two lightest quarks ($2U_\mu^{\rm(q)}$) as well as proton and
neutron ($2U_\mu^{\rm(n)}$). The $\mu$--independent part of the potential is  
thus given by $U_0+U_\om$
with $U_\om$ the $\om$--dependent contribution. For $T=0$ we obtain
\begin{eqnarray}
  \ds{U_\mu} &=& \ds{
    U_\mu^{\rm(q)}+U_\mu^{\rm(n)}
    \label{AAA171}
    } \; , \\[2mm]
  \ds{U_\mu^{\rm(q)}(\si,\om;\mu,0)} &=& \ds{
    -\frac{1}{4\pi^2}\Bigg[
    \mu_{\rm eff}\left(\mu_{\rm eff}^2-\frac{5}{2}h_q^2\si^2\right)
    \sqrt{\mu_{\rm eff}^2-h_q^2\si^2}
    }\nnn
  &+& \ds{
    \frac{3}{2}h_q^4\si^4\ln
    \frac{\mu_{\rm eff}+\sqrt{\mu_{\rm eff}^2-h_q^2\si^2}}
    {q_{\rm H}+\sqrt{h_q^2\si^2+q^2_{\rm H}}}
    \label{AAA171a}
  }\\[2mm]
  &-& \ds{
    q_{\rm H}\left(
    4q^2_{\rm H}\mu_{\rm eff}-(3q^2_{\rm H}+\frac{3}{2}h_q^2\si^2)
    \sqrt{h_q^2\si^2+q_{\rm H}^2}
    \right)\Bigg]
    \Theta(\mu_{\rm eff}^2-h_q^2\si^2-q^2_{\rm H})} \, , \nnn
  \ds{U_\mu^{\rm(n)}(\si,\om;\mu,0)} &=& \ds{
    -\frac{27}{4\pi^2}\Bigg\{
    \Bigg[\mu_{\rm eff}(\mu_{\rm eff}^2-\frac{5}{2}h_{\rm N}^2\si^2)
    \sqrt{\mu_{\rm eff}^2-h_{\rm N}^2\si^2}
    }\nnn
  &+& \ds{
    \frac{3}{2}h_{\rm N}^4\si^4\ln\frac{\mu_{\rm eff}+\sqrt{\mu_{\rm eff}^2
        -h_{\rm N}^2\si^2}}{h_{\rm N}\abs{\si}}\Bigg]
    \Theta(\mu_{\rm eff}^2-h_{\rm N}^2\si^2)
    \Theta(h_{\rm N}^2\si^2+\frac{1}{9}
    q^2_{\rm H}-\mu_{\rm eff}^2)
    }\nnn
  &+&\ds{
    \Bigg[ q_{\rm H}\left(
    \frac{4}{27}q_{\rm H}^2\mu_{\rm eff}-
    (\frac{1}{9}q_{\rm H}^2+\frac{1}{2}h_{\rm N}^2\si^2)
      \sqrt{h_{\rm N}^2\si^2+\frac{1}{9}q^2_{\rm H}} \right)
    }\nnn
  &+& \ds{
    \frac{3}{2}h_{\rm N}^4\si^4\ln\frac{q_{\rm H}+
      \sqrt{q^2_{\rm H}+9h_{\rm N}^2\si^2}}
    {3h_{\rm N}\abs{\si}}\Bigg]
    \Theta(\mu_{\rm eff}^2-h_{\rm N}^2\si^2-\frac{1}{9}q_{\rm H}^2)\Bigg\}
    \label{AAA171b}
    }\; .
\end{eqnarray}
In this expression the dependence on $\mu$ and $\omega$ appears
only implicitly through $\mu_{\rm eff}$. For given $h_q$ and $h_N$
the $\sigma$-dependence of $U_\mu$ is uniquely determined once
the dependence of $q_H$ on $\sigma$ is fixed.

The qualitative dependence of $q_{\rm H}(\si)$ on $\si$ can be inferred
from the following argument: A crucial ingredient for the confinement of
quarks in hadrons is the formation of QCD strings. Strings break because of
pair production of mesons if typical quark kinetic energies become too
large.  Therefore baryons can only exist for sufficiently small average
quark kinetic energies or momenta. Very roughly, the relevant critical
kinetic energy is expected to be proportional to the pion mass
$\sqrt{q_{\rm H}^2(\si)+h^2\si^2}\simeq2m_\pi(\si)$. The $\si$--dependence
of the pion mass can be inferred from the effective potential as
$m_\pi^2(\si)=(\prl U/\prl\si+2m_\pi^2 f_\pi)/(4\si)$, with $m_\pi$ the
pion mass in the vacuum and $f_\pi$ the pion decay constant. Since
$m_\pi^2(\si)$ always tends to zero for small enough $\si$ there should be
a critical value $\si_{\rm c}$ for which $q_{\rm H}(\si_{\rm c})=0$. We
will not use baryons for $\si<\si_{\rm c}$ and take $q_{\rm H}(\si<\si_{\rm
  c})=0$.  For our purpose we will be satisfied with a crude
approximation\footnote{For our choice $\prl q_{\rm H}(\si)/\prl\si$ is
  continuous at $\si=\si_{\rm H}$.} where we neglect the $\si$--dependence
of $q_{\rm H}$ in the range of $\si$ relevant for nuclear physics,
$\si>\si_{\rm H}$
\begin{equation}
  \label{eq:jjj001}
  q_{\rm H}(\si)=\frac{q_{\rm H}}{\si_{\rm H}^2-\si_{\rm c}^2}
  \sqrt{\left(\si^2-\si_{\rm c}^2\right)
    \left(2\si_{\rm H}^2-\si_{\rm c}^2-\si^2\right)}
    \Theta(\si-\si_{\rm c})
    \Theta(\si_{\rm H}-\si)+
    q_{\rm H}\Theta(\si-\si_{\rm H})\; .
\end{equation}
For a wide range of $\si_{\rm c}$ and $\si_{\rm H}$ our results for the
nuclear
gas--liquid transition will turn out to be independent of the precise
values of these two quantities. For definiteness we take $\si_{\rm
  c}=15\MeV$, $\si_{\rm H}=25\MeV$.  We expect that the constant $q_{\rm
  H}$ should have the size of a typical QCD scale, i.e., around $200\MeV$.
On the other hand, we will find that the quantitative aspects of the
quark--hadron transition depend on $\si_{\rm c}$, $\si_{\rm H}$ and $q_{\rm
  H}$ which parameterize in our crude approximation the effects of
confinement. This underlines that a more quantitative understanding
of the effective transition from quarks to nucleons is needed before
reliable statements about the hadron-quark phase transition at low
temperature can be made.

It is interesting to note that for  two light quark flavors ($N_{\rm c}
N_{\rm F}=6$) the $\mu$--dependent contribution to the potential at the
origin and therefore to the energy density reads for arbitrary finite
$h(\si,\mu)$
\begin{equation}
  \label{AAA013}
  \eps_\mu^{(0)}=-6U_\mu(\si=0,\om;\mu,0)=
  \frac{3}{2\pi^2}\mu_{\rm eff}^4=
  \left(\frac{3}{2}\right)^{7/3}\pi^{2/3}
  \left(n^{(0)}\right)^{4/3}\; .
\end{equation}
This has the simple interpretation of the total energy of six massless quarks
with all energy levels filled up to the Fermi energy $\mu_{\rm eff}$.
Furthermore, for $\si$ and $\mu$ in the range relevant for nuclear physics and
for sufficiently large $q_{\rm H}$, i.e.  $q_{\rm H}^2>9(\mu_{\rm
  eff}^2-h_{\rm N}^2\si^2)$, the contribution $U_\mu^{\rm(n)}$ is simply the
mean field result for a nucleon meson model, whereas $U_\mu^{\rm(q)}$
vanishes.  Our approach gives a new motivation for the approximate validity of
mean field theory from the truncation of an exact flow equation. Furthermore,
it offers the possibility of a systematic improvement, e.g., by taking the
$\mu$--dependence of $h_{\rm N}$ into account.  Despite this similarity, our
method goes beyond mean field theory in an important aspect: For the free
energy only the difference between vanishing and non--vanishing chemical
potential is described by mean field theory, whereas
we do not rely on mean field results for the effective action
$\Gamma$ at $\mu=0$.  Since $\Gamma(\mu=0)$ is the
generating functional for the propagators and vertices in vacuum it can, in
principle, be directly related to measured properties like meson masses and
decays. This is very important in practice,
since mean field theory does not give a very
reliable description of the vacuum properties.

\section{Meson interactions}
\label{sec:04}

In order to discuss possible phase transitions as $\mu$ is increased beyond a
critical value we need information about the effective potential for $\mu=0$.
For a vacuum without spontaneous symmetry breaking relatively accurate
information about $U_0(\si;T)=U_0(\si,\om=0;\mu=0,T)$ for all relevant $\si$
could be extracted from the knowledge of meson masses and interactions.  Also
the approximation~(\ref{AAA006}) for the fermionic propagator would presumably
be reasonable for arbitrary $\si$.  In case of spontaneous chiral symmetry
breaking the situation is more complex: The true
effective potential $U_0$ becomes
convex because of fluctuations which interpolate between the minima of the
``perturbative'' or ``coarse grained'' potential~\cite{RW90-1,TW92-1}.  Masses
and interactions give only information about the ``outer region'' of the
potential which is not affected by this type of fluctuations. In parallel, the
simple form of the fermionic propagator~(\ref{AAA006}) becomes invalid in the
``inner region'' for small $\si$ because of a complex momentum
dependence~\cite{RW90-1,TW92-1} and the breakdown of the approximation of a
constant Yukawa coupling.
In order to cope with these difficulties, $U_0(\si;T)$ should rather be
associated with a coarse grained effective potential. For a suitable coarse
graining scale\footnote{The coarse graining scale $k$ is chosen such that
$U_k$ is approximately $k$--independent for $\abs{\si}$ around $\abs{\olsi}$
or larger, whereas the approach to convexity for $\abs{\si}<\abs{\olsi}$ and
$k\ra0$ has not yet set in.} $k$ the effect of the omitted fluctuations with
momenta smaller than $k$ is expected to be small near the $\mu$--dependent
minimum of $U$.  Around the minimum at $\sigma_0$
we can therefore continue to associate
$U_0(\si;T)$ with the effective potential and relate its properties to the
measured masses and decay constants. On the other hand, we do not have much
information about the shape of $U_0(\si;T)$ for $\si\simeq0$.  This
uncertainty in the appropriate choice of $U_0(\si;T)$ is one of the main
shortcomings of our method. In practice, we interpolate the partly known
polynomial form of $U_0(\si;T)$ form the outer region (which includes the
minimum characterizing the vacuum) to the inner region for small $\si$.  By
continuity, this should be quite reasonable for nuclear matter since the
relevant values of $\si$ are not much smaller than the vacuum expectation
value $\si_0$.  For quark matter, the uncertainties are more important.

We investigate here the two flavor case with a potential of the form
\begin{eqnarray}
  \label{AAA030}
  \ds{U_0(\si;T)} &\equiv& \ds{
    2m_\pi^2(T)\left[\si^2-\si_0^2(T)\right]+
    2\la(T)\left[\si^2-\si_0^2(T)\right]^2+
    }\nnn
  &+& \ds{
    \frac{4}{3}\frac{\gm_3(T)}{\si_0^2(T)}
    \left[\si^2-\si_0^2(T)\right]^3+
    \frac{\gm_4(T)}{\si_0^4(T)}
    \left[\si^2-\si_0^2(T)\right]^4
      }\nnn
  &+& \ds{
    \frac{4}{5}\frac{\gm_5(T)}{\si_0^6(T)}
    \left[\si^2-\si_0^2(T)\right]^5-
      2\jmath\si+c(\jmath,T)
      }
\end{eqnarray}
where
\begin{equation}
  \jmath= 2m^2_\pi(0)\si_0(0)
  \; ,\;\;\;
  c(\jmath,0)=2\jmath\si_0(0)
  \; .
\end{equation}
In the remainder of this work we mainly consider $T=0$ and use $\lambda\equiv
\lambda(0),\ U(\si;\mu)\equiv U(\si;\mu,0)$ etc.  The meson field is
normalized such that $\si_0=\si_0(0)$ is related to the pion decay constant by
$\si_0=f_\pi/2=46.5\MeV$.  This means that the pions have a standard kinetic
term (as derived from $\Lc_{\rm kin}^{(0)}=\Tr\prl_\mu\Phid\prl^\mu\Phi$).
Because of higher order kinetic invariants~\cite{JW98-1} the kinetic term for
the sigma meson, $\Lc_{\rm kin,\si}=2Z_\si\prl_\mu\si\prl^\mu\si$ can involve
a wave function renormalization $Z_\si$ different from the one for the pions.
The potential~(\ref{AAA030}) arises from a fifth order polynomial in the
invariant $\rho= {\rm Tr}\Phid\Phi=2\si^2$ with an additional source term
$-\frac{1} {2}{\rm Tr}\jmath(\Phi+\Phid)$, where $\jmath$ is proportional to
the renormalized current quark mass (say at 1\GeV). The only violation of the
chiral $SU_L(2)\times SU_R(2)$ symmetry arises from this source and in the
chiral limit of vanishing current quark masses the last two terms in
eq.~(\ref{AAA030}) should be dropped. The coupling $\la$ is related to the
$\si$--mass $m_\si$ by $\tilde{m}_\si^2=Z_\si m_\si^2= m_\pi^2+4\la\si_0^2$.
We will use here $\tilde{m}_\si=510\MeV$, $\la=28$.  It is actually
$\tilde{m}_\si$ rather than the physical mass $m_\si$ which is relevant
for the properties of nuclear matter.  One of the parameters $\gm_3$,
$\gamma_4$ or $\gm_5$ can be eliminated in favor of the scale $\mu_0$ which
characterizes the hight of $U_0$ at the origin
\begin{equation}
  \label{AAA031}
  \frac{\mu_0^4}{2\pi^2}\equiv
  U(0;0)=\si_0^4\left(2\la-\frac{4}{3}\gm_3+\gm_4-
\frac{4}{5}\gm_5\right)+
  2m_\pi^2\si_0^2\; .
\end{equation}
Without the complications of confinement (i.e., for $q_{\rm H}=0$) the
quark--hadron phase transition in the chiral limit ($\jmath=0$) would occur
for $\mu_{\rm eff}=\mu_0$.

Finally, we determine the expectation value of $\om$ by observing the
identity
\begin{equation}
  \label{eq:AAA007}
  \frac{\prl U_\mu}{\prl\om}=
  \frac{2}{3}g_{\om}
  \frac{\prl U_\mu}{\prl\mu}\; .
\end{equation}
It follows from first differentiating eqs.~(\ref{AAA011}), (\ref{AAA014a})
with respect to $\om$ and then performing the $\mu$--integration. For the
$\mu=0$ contribution we only take into account a $\si$-- and $T$--dependent
mass term
\begin{equation}
  \label{eq:AAA008}
  U_\om=-\frac{1}{2}M_\om^2(\si,T)\om^2\; .
\end{equation}
The solution of the $\om$--field equations for arbitrary $\si$, $T$, $\mu$
obeys
\begin{equation}
  \label{eq:AAA009}
  \ol{\om}(\si,\mu,T)=
  \frac{2g_{\om}}{3 M_\om^2}
  \frac{\prl U_\mu}{\prl\mu}(\si,\ol{\om};\mu,T\; .)
\end{equation}
We note that at the potential minimum $\ol{\om}$ is proportional to the
baryon density with a negative coefficient. This implies
that the coupling to $\om$ reduces the effective chemical
potential~\cite{Kle92-1}. In the following we will always assume that
$\ol{\om}(\si,\mu,T)$ is inserted such that $\mu_{\rm eff}$ becomes a
function of $\si$, $\mu$ and $T$. The field equation which determines
the location $\olsi$ of the minimum of the effective potential
can be expressed in terms of partial $\si$--derivatives at fixed
$\mu_{\rm eff}$
\begin{equation}
  \label{eq:AAA009a}
  \frac{\prl U_0}{\prl\si}(\olsi)-
  M_\om(\olsi)
  \frac{\prl M_\om}{\prl\si}(\olsi)\ol{\om}^2+
  2\frac{\prl U_\mu}{\prl\si}_{\big|_{\ds{\mu_{\rm eff}}}}(\olsi)=0\; .
\end{equation}
Below we will also neglect a possible $\sigma$-dependence of $M_\omega$.
The location of the minimum $\ol{\sigma}$ becomes then independent
of the value of $\ol{\om}$ and only depends on $\mu_{\rm eff}$. 
For this setting the coupling to vector mesons
is relevant only for the relation between $\mu_{\rm eff}$ and $\mu$.

\section{Meson--baryon interactions \label{mesbar}}

A crucial ingredient for any quantitative analysis is the
sigma--nucleon coupling $h_{\rm N}$. We first investigate if chiral
symmetry and the observed value of the pion nucleon coupling place any
restrictions on this coupling. For this purpose we employ a derivative
expansion of the most general effective Lagrangian which is bilinear in the
nucleon doublet field $\Psi_{\rm N}$ and involves scalar and pseudoscalar
fields contained in the $2\times2$ matrix $\Phi$
\begin{eqnarray}
  \label{eq:AAA009b}
  \ds{\Lc} &=& \ds{
    \frac{1}{2}\Bigg\{\ol{\Psi}_{\rm N R}
    F(\Phi\Phid,\rho)\Phi\Psi_{\rm N L}-
    \ol{\Psi}_{\rm N L}F(\Phid\Phi,\rho)\Phid\Psi_{\rm N R}
    }\nnn
  &+& \ds{
    \ol{\Psi}_{\rm N L}G_1(\Phi\Phid,\rho)
    i\gm^\mu\prl_\mu\Psi_{\rm N L}+
    \ol{\Psi}_{\rm N R}
    G_1(\Phid\Phi,\rho)
    i\gm^\mu\prl_\mu\Psi_{\rm N R}
    }\\[2mm]
  &+& \ds{
    \ol{\Psi}_{\rm N L}\Phid
    G_2(\Phi\Phid,\rho)i\gm^\mu
    \left(\prl_\mu\Phi\right)\Psi_{\rm N L}+
    \ol{\Psi}_{\rm N R}\Phi
    G_2(\Phid\Phi,\rho)i\gm^\mu
    \left(\prl_\mu\Phid\right)\Psi_{\rm N R}+
    {\rm  h.c.\Bigg\}}
    }\nonumber\; .
\end{eqnarray}
Here we have imposed $\Pc$ and $\Cc$ symmetry and used $\Psi_{\rm N
  L}=(1+\gm_5)\Psi_{\rm N}/2$. With the standard decomposition
\begin{equation}
  \label{AAA009c}
  \begin{array}{rclcrcl}
    \ds{\Phi} &=& \ds{
      \si\xi^2=\si U} &,& \ds{\xi} &=& \ds{
      \exp\left(\frac{i}{4\si}\vec{\tau}\vec{\pi}\right)
      }\nnn
    \ds{N_{\rm L}} &=& \ds{\xi\Psi_{\rm N L}} &,&
    \ds{N_{\rm R}} &=& \ds{\xi^\dagger\Psi_{\rm N R}
      }
  \end{array}
\end{equation}
one finds
\begin{eqnarray}
  \label{AAA009d}
  \ds{\Lc} &=& \ds{
    3h_{\rm N}(\si)\si\ol{N}\gm_5 N+
    Z_{\rm N}(\si)\ol{N}
    \left(i\gm^\mu\prl_\mu-\gm^\mu v_\mu+
      G_{\rm A}(\si)\gm^\mu\gm^5 a_\mu\right) N
      }\nnn
    &-& \ds{i
      \frac{Z_{\rm N}(\si)}{2\si^2}
      \left[ G_{\rm A}(\si)-1\right]
        \si(\prl_\mu\si)
        \ol{N}\gm^\mu N}
\end{eqnarray}
where
\begin{equation}
  \label{AAA009dd}
  \begin{array}{rclcrcl}
    h_{\rm N}(\si) &=& F(\si^2,2\si^2)/3 &,&
    Z_{\rm N}(\si) &=& G_1(\si^2,2\si^2)\nnn
    G_{\rm A}(\si) &=& \ds{1-
      \frac{2G_2(\si^2,2\si^2)\si^2}
      {G_1(\si^2,2\si^2)} }
  \end{array}
\end{equation}
and
\begin{eqnarray}
  \label{AAA009e}
  \ds{v_\mu} &=& \ds{
    -\frac{i}{2}
    \left(\xi^\dagger\prl_\mu\xi+
      \xi\prl_\mu\xi^\dagger\right)
      }\nnn
  \ds{a_\mu} &=& \ds{
    -\frac{i}{2}
    \left(\xi^\dagger\prl_\mu\xi-
      \xi\prl_\mu\xi^\dagger\right)=
      \frac{1}{4\si}\vec{\tau}\prl_\mu\vec{\pi}+\ldots
      }\; .
\end{eqnarray}
Normalization of the baryon number current requires $Z_{\rm N}(\si_0)=1$
and we neglect the $\si$--dependence of $Z_{\rm N}$ in the following. The
strength of the linear pion--nucleon coupling is fixed by $g_{\rm A}=G_{\rm
  A}(\si_0)$ and bares no relation to the function $h_{\rm N}(\si)$.
We may expand $h_{\rm N}(\si)$ around $\si_0$
\begin{equation}
  \label{AAA009f}
  h_{\rm N}(\si)=h_{\rm N}(\si_0)+
  \frac{g_{\rm N}}{\si_0^2}
  \left(\si^2-\si_0^2\right)+\ldots\; .
\end{equation}
With $m_{\rm N}=3h_{\rm N}(\si_0)\si_0=939\MeV$ we find $h_{\rm
  N}(\si_0)=6.73$. Linearizing $m_{\rm N}(\si)=3h_{\rm N}(\si)\si$ around
$\si_0$ then yields $m_{\rm N}(\si)=3\tilde{h}\si+\eps_{\rm G}$ with
$\tilde{h}=h_{\rm N}(\si_0)+2g_{\rm N}$, $\eps_{\rm G}=-6g_{\rm N}\si_0$.
The linear sigma--nucleon coupling $\tilde{h}$ is a free parameter which is
expected to be in the vicinity of $h_{\rm N}(\si_0)$. We will determine it
below from the properties of nuclear matter. Since $\tilde{h}$ also appears
in the scattering of nucleons a comparison with experiment may serve as a
test for our model.

\section{Nuclear matter and the nuclear phase transition}
\label{sec:06}

\begin{center}
  \begin{figure}[htb]
    \unitlength1.0cm
    \begin{picture}(9.,7.)
      \put(0.6,5.7){$\ds{\frac{U(\si,\mu_{\rm nuc})}{\MeV^4}}$}
      \put(8.0,-0.2){$\si/\MeV$}
      \put(-0.5,-16.8){
        \epsfysize=26.cm
        \epsffile{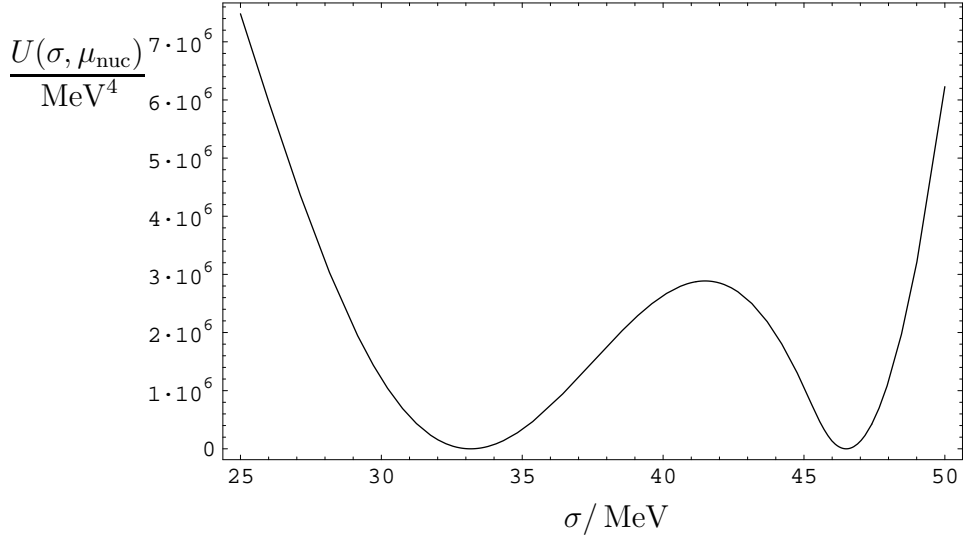}
        }
    \end{picture}
    \caption{\footnotesize The critical effective potential for the nuclear
      liquid--gas transition corresponding to the parameter set A of
      table~\ref{tab:01}.}
    \label{nucl_pot_crit}
  \end{figure}
\end{center}

Let us turn to the zero temperature properties of nuclear matter in our
picture. For $\mu=0$ the effective potential or free energy $U$ has its
minimum at $\sigma_0=f_\pi/2=46.5\MeV$.  The potential in the region near
$\si_0$ is not altered as long as $\mu_{\rm eff}$ remains small enough
(cf.~eq.~(\ref{AAA014a})).  This changes
as $\mu_{\rm eff}$ is increased beyond a critical
threshold. For suitable parameters in $U_0$
(eq. (\ref{AAA030})) we observe that for $3 \mu$ somewhat below the
nucleon mass a new minimum of $U$ occurs at
$\ol{\si}^{\rm(nuc)}(\mu)<\si_0$, with a potential barrier between both
minima.  For a certain range of $\mu$ the local minimum at
$\ol{\si}^{\rm(nuc)}(\mu)$ and the global minimum at $\si_0$ coexist.  As
$\mu$ increases, the value of $U(\ol{\si}^{\rm(nuc)}(\mu))$ is lowered
whereas $U(\si_0)=0$ remains fixed as long as the effective chemical
potential is smaller than a third of the nucleon mass, $\mu_{\rm
  eff}<h_{\rm N}(\si_0)\si_0$.  There is a critical value $\mu_{\rm nuc}$
for which the two minima at
\begin{equation}
  \label{eq:rfrf01}
  \olsinuc\equiv\ol{\si}^{\rm(nuc)}(\mu_{\rm nuc})
\end{equation}
and $\si_0$ are
degenerate, $U(\olsinuc,\mu_{\rm nuc})=U(\si_0,\mu_{\rm nuc})=0$.  The
corresponding critical potential is plotted in figure~\ref{nucl_pot_crit}.
Both phases have equal, vanishing pressure $p=-U$ and can coexist.  We observe
that the phase transition between the vacuum $(\si=\si_0)$ and nuclear matter
$(\si=\olsinuc)$ is clearly of first order.  For small temperature this
corresponds to the transition between a gas of nucleons and nuclear matter
which may be associated with a nuclear liquid.

For a quantitative description we concentrate mainly on large values of
$q_{\rm H}$ where this transition happens in a region with $\mu_{\rm
  eff}^2< h_{\rm N}^2(\olsinuc)\olsinuc^2+q_{\rm H}^2/9$.  In this case the
transition from nucleon to quark degrees of freedom does not affect nuclear
properties and the phase transition from a nucleon gas to nuclear matter.
We partly recover the $\si$--$\om$--model of nuclear
physics~\cite{SW97-1,Wal95-1,Gle97-1}, in a context where chiral symmetry
breaking and constraints from meson masses and decays are properly
incorporated. For large enough $q_{\rm H}$ the critical baryon density of
the nuclear liquid is given by eq.~(\ref{AAA014a})
\begin{equation}
  \label{eq:AAA011}
  n_{\rm nuc}=\frac{18}{\pi^2}
  \left[\mu_{\rm eff}^2(\mu_{\rm nuc},\olsinuc)-
    h_{\rm N}^2(\olsinuc)\olsinuc^2\right]^{3/2}\; .
\end{equation}
We will see below that one can identify $n_{\rm nuc}$ with the baryon density
in nuclei $\overline{n}^{(n)}=1.175 \times 10^6\cdot\MeV^3$ up to small
corrections. Furthermore, the baryon number independent contribution to the
binding energy per nucleon in a large sample of nuclear matter is known from
the mass formula for nuclei: $\bt^{\rm(n)}=-16.3\MeV$. In our context one
finds $\bt^{\rm(n)}=3\mu_{\rm nuc}-m_{\rm N}$ and for realistic models the
gas--liquid transition should therefore occur for $\mu_{\rm nuc}=307.57\MeV$.
Eq.~(\ref{eq:AAA011}) then yields a quantitative relation between the
effective chemical potential in nuclear matter $\mu_{\rm eff,nuc}=\mu_{\rm
  eff} (\mu_{\rm nuc},\olsinuc)$ and the effective nucleon mass $m_{\rm
  N}(\olsinuc)=3h_{\rm N}(\olsinuc)\olsinuc=\eps_{\rm G}+3\tilde h\olsinuc$.
For $m_{\rm N}(\olsinuc)/m_{\rm N}=(0.6,0.65,0.7,0.75,0.8)$ one finds
$\mu_{\rm eff,nuc} =(206.6,220.9,235.4,250.1,264.8)\MeV$.  Equivalently, this
can be seen as a relation between $h_{\rm N}(\olsinuc)\olsinuc$ and the
coupling $g^{(\om)}$ if we use with (\ref{eq:AAA009a})
\begin{equation}
  \label{eq:fff001}
  \mu_{\rm eff,nuc}=\mu_{\rm nuc}-\frac{1}{3}
  \frac{g_\om^{2}}{M_\om^2}n_{\rm nuc}\; .
\end{equation}
For a $\si$--independent $\om$--mass $M_\om=783\MeV$ typical values for the
above ratios for $m_{\rm N}(\olsinuc)/m_{\rm N}$ are
$g^{(\om)}=(12.61,11.68,10.66,9.52,8.21)$.  From the value of nuclear density
we can compute the Fermi momentum $q_{\rm nuc}=259\MeV$. This yields for this
scenario a lower bound $q_{\rm H}> q_{\rm nuc}= 259\MeV$.  (For quantitative
computations we take $q_{\rm H}=1.2\,q_{\rm nuc}$.)
\begin{table}[htb]
  \begin{center}
    \leavevmode
    \begin{tabular}{|c||c|c|c|c|c|c|}
      \hline
      &
      $\tilde{h}$ &
      $g^{(\om)}$ &
      $\gm_3$ &
      $\gm_4$ &
      $\gm_5$ &
      $\frac{\mu_0}{\MeV}$
      \\[1mm] \hline\hline
      $A$ &
      $5.4$ &
      $9.02$ &
      $-30$ &
      $47$ &
      $-60$ &
      $372$
      \\\hline
      $B$ &
      $5.0$ &
      $9.52$ &
      $19$ &
      $112$ &
      $0$ &
      $348$
      \\\hline
      $C$ &
      $4.6$ &
      $8.74$ &
      $0$ &
      $55$ &
      $0$ &
      $330$
      \\\hline
    \end{tabular}
    \caption{\footnotesize Coupling constants for three different parameter
      sets. The linear sigma--nucleon coupling $\tilde{h}$, the coupling
      $g^{(\om)}$ of the $\om$--meson to the $u,d$--quarks and the nucleons,
      the meson self--interactions $\gm_3$, $\gm_4$, $\gm_5$ and the scale
      $\mu_0$ are defined in sections $\ref{sec:04}$ and $\ref{mesbar}$.}
    \label{tab:01}
  \end{center}
\end{table}

An important quantity for the equation of state is the compression modulus
\begin{eqnarray}
  \label{fff001a}
  \ds{K} &=& \ds{
    9n^2\frac{d^2}{d n^2}
    \left(\frac{\eps}{n}\right)=
    9\left(\frac{d p}{d n}-2\frac{p}{n}\right)\; .
    }
\end{eqnarray}
The value at the phase transition
\begin{eqnarray}
  \ds{K_0} &=& \ds{
    9\left.\frac{d p}{d n}\right|_{n_{\rm nuc}}=
    27n_{\rm nuc}
    \left.\frac{d\mu}{d n}\right|_{n_{\rm nuc}}
    }\nnn
  &=& \ds{
    9n_{\rm nuc}\frac{g^{(\om)2}}{M_\om^2}+
    \frac{1}{\mu_{\rm eff,nuc}}
    \left(\frac{3\pi^2 n_{\rm nuc}}{2}\right)^{2/3}
    }\nnn
  &+& \ds{
    9n_{\rm nuc}\frac{d\olsinuc}{d n_{\rm nuc}}
    \left[
      \frac{\tilde{h}m_{\rm N}(\olsinuc)}
      {\mu_{\rm eff,nuc}}+
      n_{\rm nuc}\frac{d}{d\olsi}
      \left(\frac{g^{(\om)2}}{M_\om^2}\right)
      \right]
      \label{fff001aa}
    }
\end{eqnarray}
has been inferred from experiment as
$K_0=(210-220)\MeV$~\cite{BBDG95-1,Gle97-1}. This can be used to obtain
additional information about the $\mu$--independent
part $U_0$ of the effective
potential. Neglecting the $\si$--dependence of $M_\om$ one finds by
differentiating eq.~(\ref{eq:AAA009a})
\begin{eqnarray}
  \label{fff001b}
  \ds{\frac{d\olsi}{d n}} &=& \ds{
    -\frac{\tilde{h}m_{\rm N}(\olsi)}{\mu_{\rm eff}}
    \Bigg[\frac{\prl^2 U_0}{\prl\si^2}(\olsi)+
    2\frac{\prl^2 U_\mu}{\prl\si^2}_{\big|_{\ds{\mu_{\rm eff}}}}(\olsi)+
    \frac{6}{\pi^2}\tilde{h}^2m_{\rm N}^2(\olsi)
    \frac{\sqrt{9\mu_{\rm eff}^2-m_{\rm N}^2(\olsi)}}
    {\mu_{\rm eff}}\Bigg]^{-1}
    }\; .
\end{eqnarray}
Combining eqs.~(\ref{fff001b}) and~(\ref{fff001aa}) the compression modulus
yields information about $\frac{\prl^2 U_0}{\prl\si^2}(\olsinuc)$ in
addition to $U_0(\olsinuc)$ and $\frac{\prl U_0}{\prl\si}(\olsinuc)$ which
are determined (for given $m_{\rm N}(\olsinuc)$ and $\tilde{h}$) by the
condition $U(\olsi)=0$ and the field equation~(\ref{eq:AAA009a}).

For any given value of
the coupling $\gm_5$ the system of equations provides a mapping
between the parameters $(\tilde{h},g^{(\om)},\gm_3,\gm_4)$ and the
quantities $(n_{\rm nuc},\bt,K_0,m_{\rm N}(\olsinuc))$. For a demonstration
of the range of values for various quantities of interest we report our
results for two parameter sets with different $\gm_5$ (A and B) in
tables~\ref{tab:01}--\ref{tab:03}. (For both sets $\bt=-16.3\MeV$ and
$n_{\rm nuc}=\ol{n}^{\rm(n)}$.) Agreement with nuclear properties
can indeed be achieved. It is not our aim here to make a precise
determination of parameters and we only mention that somewhat smaller
values of $m_N(\ol{\sigma}_{\rm nuc})$ or other (large) values of the
compression modulus lead to qualitatively similar results. One finding
remains common, however: for nuclear matter properties in a reasonable
range we always need large values of some of the couplings
$|\gamma_3|,|\gamma_4|$ or $|\gamma_5|$. No viable solution
was found for an approximately quartic meson potential
with small $|\gamma_{3,4,5}|$. This may be a cause of worry for this
class of models since earlier renormalization group studies of the
effective meson potential have typically resulted in substantially
smaller higher order couplings $|\gamma_{3,4,5}|$ than the ones
needed here \cite{BJW98-1}.

For given parameters we can also compute the energy density and
the pressure and relate it to the baryon density. This determination
of the equation of state of dense nuclear matter is rather
insensitive to aspects of QCD not treated here like the role of
gluons or vector mesons (beyond the effect of $\ol{\omega}\not=0$).
The reason is that these degrees of freedom do not contribute to the
difference of the effective action between zero and nonzero
chemical potential. The nuclear equation of state can therefore
be considered as a prediction of the model (for fixed parameters).

\begin{table}[htbp]
  \begin{center}
    \leavevmode
    \begin{tabular}{|c||c|c|c|c|c|}
      \hline
       &
      $\frac{\olsinuc}{\MeV}$ &
      $\frac{m_{\rm N}(\olsinuc)}{m_{\rm N}}$ &
      $\frac{K_0}{\MeV}$ &
      $\frac{\Si}{\Si^{\rm(n)}}$ &
      $\frac{\mu_{\rm eff,nuc}}{\MeV}$
      \\[1mm] \hline\hline
      $A$ &
      $33.2$ &
      $0.77$ &
      $214$ &
      $1$ &
      $256$
      \\\hline
      $B$ &
      $30.85$ &
      $0.75$ &
      $217$ &
      $1.3$ &
      $250$
      \\\hline
      $C$ &
      $29.8$ &
      $0.755$ &
      $\infty$ &
      $2$ &
      $259$
      \\\hline
    \end{tabular}
    \caption{\footnotesize Properties of nuclear matter at vanishing
      pressure for the parameters of table~\ref{tab:01}. The table shows
      values for the chiral order parameter $\olsinuc$, the effective nucleon
      mass $m_{\rm N}(\olsinuc)$, the compression modulus $K_0$ and the
      effective chemical potential $\mu_{\rm eff,nuc}$ for nuclear matter. The
      surface tension $\Si^{\rm}$ for the droplet model of nuclei normalized
      to the value $\Si^{\rm(n)}$ extracted from the nuclear mass formula is
      discussed in section \ref{dropmod}.}
    \label{tab:02}
  \end{center}
\end{table}

\begin{center}
  \begin{figure}[htb]
    \unitlength1.0cm
    \begin{picture}(9.,7.)
      \put(1.4,6.0){$\ds{\frac{\bt}{\MeV}}$}
      \put(8.0,-0.2){$n/\ol{n}^{\rm (n)}$}
      \put(-0.5,-16.8){
        \epsfysize=26.cm
        \epsffile{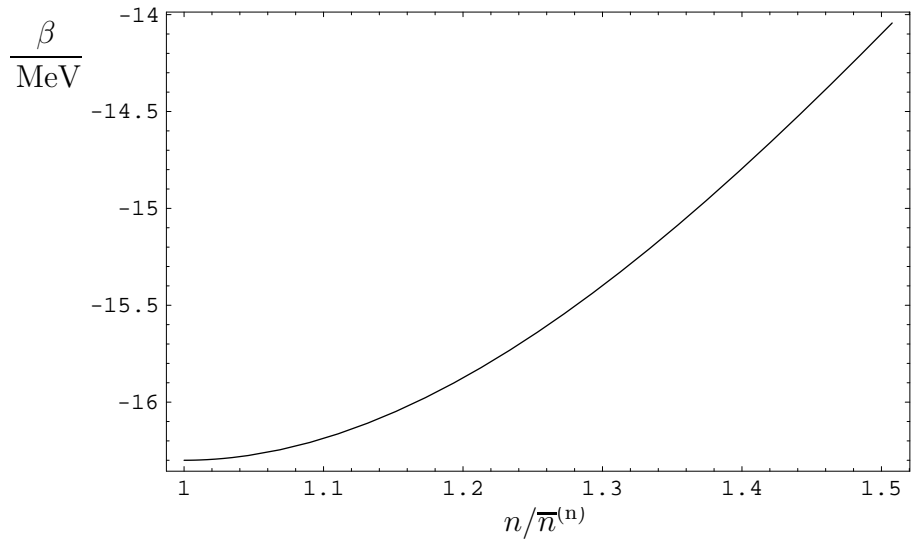}
        }
    \end{picture}
    \caption{\footnotesize Binding energy per nucleon $\bt$ as a function of
      density. Parameter values correspond to A in table~\ref{tab:01}.}
    \label{eG}
  \end{figure}
\end{center}

\begin{center}
  \begin{figure}[htb]
    \unitlength1.0cm
    \begin{picture}(9.,7.)
      \put(1.6,6.0){$\ds{\frac{n}{\ol{n}^{\rm(n)}}}$}
      \put(8.0,-0.2){$p/\MeV^4$}
      \put(-0.5,-16.8){
        \epsfysize=26.cm
        \epsffile{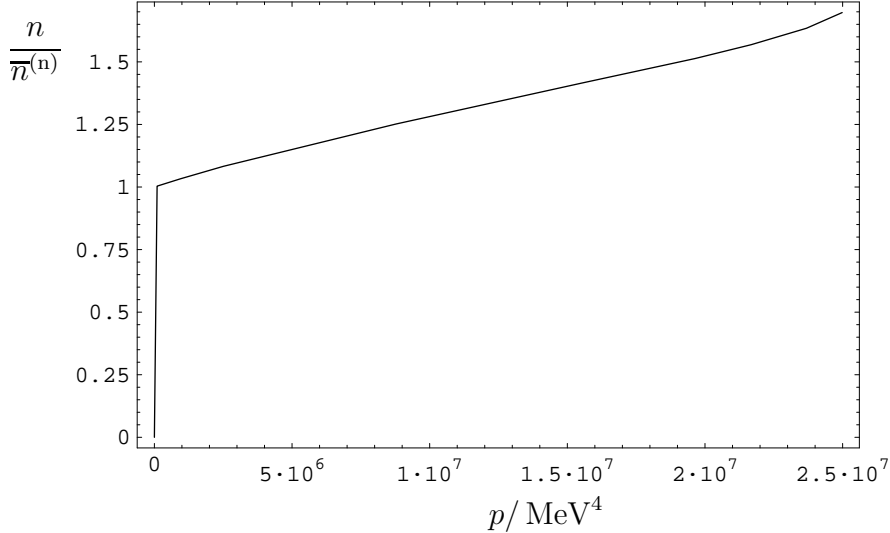}
        }
    \end{picture}
    \caption{\footnotesize Baryon density as a function of
      pressure in the vicinity of the nuclear gas--liquid transition
      at very low $T$.
      Parameters correspond to set A in table~\ref{tab:01}.}
    \label{n_p}
  \end{figure}
\end{center}

In figure~\ref{eG} we have plotted the binding energy per nucleon,
$\bt=\eps/n-m_{\rm N}$, as a function of density corresponding to the
parameter set $A$. For values of $n$ larger than approximately $1.7$ the
details of the transition from nuclear to quark degrees of freedom become
important and we don't expect our results to remain quantitatively
reliable.  Similarly, our results for the baryon density as a function of
pressure are displayed in figure~\ref{n_p}. Figures~\ref{eG}
and~\ref{n_p} can be combined to yield the equation of state $\eps(p)$ for
$n<1.5\ol{n}^{\rm(n)}$.

Within the `` $\sigma-\omega$ model'' (models A and B) the dominant
repulsion between nucleons at short distance is ascribed to the exchange
of $\omega$-mesons. It is not established if this
repulsion is indeed sufficient. Another possible repulsion mechanism
is the effective transition from nucleons to quarks at short distance.
In view of the potential difficulties of the $\sigma-\omega$ model
we also explore this second alternative --- our model C.
For this purpose
it is instructive to consider  an extreme scenario where the
characteristic quark--baryon transition momentum $q_H$ takes on its lower
bound
\begin{equation}
  \label{eq:fff002}
  q_{\rm H}(\olsinuc)=q_{\rm H}=q_{\rm nuc}=259\MeV
\end{equation}
The results correspond to the set $C$
of tables~\ref{tab:01}--\ref{tab:03}.  Because of
the $\Th$--function in eq.~(\ref{AAA014a}) the nucleon contribution to the
density does not increase any more
as $\mu_{\rm eff}$ exceeds the critical value
given by eq.~(\ref{eq:fff001}). On the other hand, there is a range of
$\mu_{\rm eff}$ for which the quark fluctuations~(\ref{AAA011}) do not yet
contribute to the baryon density.  For this range the density $n$ will not
depend on any other parameter of the model and $n_{\rm
  nuc}=\ol{n}^{\rm(n)}$ is guaranteed by eq.~(\ref{eq:fff002}). Details of
the potential in the vicinity of $\olsinuc$ are now affected by the
transition from nucleon to quark degrees of freedom.  For the simple
choice~(\ref{eq:jjj001}), however, they do not depend on $\si_c$ or
$\si_{\rm H}$ provided both are smaller than $\olsinuc$. Because of the gap
in $\mu_{\rm eff}$ between the nucleon Fermi surface and the onset of quark
fluctuations many properties become very simple. The minimum occurs within
the range $\si_{\rm q}<\olsi_{\rm nuc}<\si_{\rm nf}$. Here $\si_{\rm q}$
corresponds to the onset of quark fluctuations
\begin{equation}
  \label{AAA103}
  \si_{\rm q}(\mu)\equiv
  \frac{1}{h}\sqrt{\mu_{\rm eff}^2-q_{\rm H}^2(\si_{\rm q})}
\end{equation}
whereas $\si_{\rm nf}$ denotes the maximal value of $\si$ for which
all nucleon levels with $\vec{q}\,^2\le q_{\rm H}^2$ are filled
\begin{equation}
  \label{AAA104}
  \si_{\rm nf}(\mu)\equiv\frac{1}{h_{\rm N}(\si_{\rm nf})}
  \sqrt{\mu_{\rm eff}^2-\frac{1}{9}q_{\rm H}^2(\si_{\rm nf})}\; .
\end{equation}
For values of $\olsi^{\rm(nuc)}(\mu)$ between ${\rm max}(\si_{\rm
  q},\si_{\rm H})$ and $\si_{\rm nf}$ the baryon density is independent of
$\mu$
\begin{equation}
  \label{AAA105}
  n(\mu)=\frac{2}{3\pi^2}q_{\rm H}^3=
  \ol{n}^{\rm(n)}\; .
\end{equation}
This would give a natural explanation for a large compression
modulus $K$ according to eq. (\ref{fff001aa}).
Also, for ${\rm max}(\si_{\rm q},\si_{\rm H})<\si<\si_{\rm nf}$ one finds
that $\prl U_\mu/\prl\mu$ is independent of $\si$ and the constant
shift~(\ref{eq:fff001}) between $\mu_{\rm eff}(\si,\mu)$ and $\mu$ holds
for all $\si$.  The relation between $n_{\rm nuc}$ and $q_{\rm H}$ is such
that up to a Fermi momentum $q_{\rm H}(\olsinuc)=q_{\rm H}$ all levels are
filled with nucleons (or bound quarks).  In this crude picture the higher
momentum levels (corresponding to a larger baryon number in a fixed volume)
would have to be filled by free (constituent) quarks. This leads to a
particularly simple explanation why nuclear density is almost independent
of all other parameters characterizing the state of nuclear matter at
$T=0$, like pressure, baryon number or the $Z/B$ ratio of a nucleus.
Typical parameter values and corresponding characteristics of nuclear
matter for this ``saturation scenario'' can be found  as set C
in  tables 1 and 2.

We next discuss the equation of state for the saturation scenario.
For a given value of $\si$ one finds in this scenario a range $\mu_{\rm
  nf}<\mu_{\rm eff}<\mu_{\rm q}$ with constant $\prl U_\mu/\prl\mu$ where
\begin{eqnarray}
  \label{AAA106}
  \ds{\mu_{\rm nf}(\si)} &=& \ds{
    \sqrt{h_{\rm N}^2(\si)\si^2+\frac{1}{9}q_{\rm H}^2(\si)}
    }\nnn
  \ds{\mu_{\rm q}(\si)} &=& \ds{
    \sqrt{h_q^2\si^2+q_{\rm H}^2(\si)}
    }\; .
\end{eqnarray}
In this range $\ol{\om}$ is independent of $\mu$ and $U_\mu$ has the simple
form
\begin{equation}
  \label{AAA107}
  U_\mu(\si;\mu)=U_\mu(\si;\ol{\mu})-
  \frac{1}{\pi^2}\left(\mu-\ol{\mu}\right) q_{\rm H}^3(\si)\; .
\end{equation}
Here $\ol{\mu}$ is a fixed reference value within the interval $[\mu_{\rm
  nf},\mu_{\rm q}]$, and we remind the reader that $q_{\rm H}(\si>\si_{\rm
  H})=q_{\rm H}$.  We note that the location of the potential minimum at
$\ol{\si}^{\rm(nuc)}(\mu)$ is independent of $\mu$. Using
$\ol{\mu}=\mu_{\rm nuc}$, the pressure and energy density of nuclear matter
are
\begin{eqnarray}
  \label{AAA114a}
  \ds{p} &=& \ds{
    2\frac{q_{\rm H}^3}{\pi^2}
    \left(\mu-\mu_{\rm nuc}\right) }\\[2mm]
  \label{AAA114b}
  \ds{\eps} &=& \ds{
    \frac{2}{\pi^2}q_{\rm H}^3\mu-
    p=
    \left( m_{\rm N}+\bt\right) n
    }\; .
\end{eqnarray}
The ($T=0$) equation of state for nuclear
matter for this extreme saturation scenario
\begin{equation}
  \label{AAA117}
  \frac{\prl\eps}{\prl p}=0\;,\;\;\;
  \frac{\prl n}{\prl p}=0\; ,
\end{equation}
implies a diverging compression modulus $K_0$ and is therefore not fully
realistic. Nevertheless, it is well conceivable that the true behavior of
nuclear matter is somewhere between the simple version of the $\si$--$\om$
model and the extreme saturation scenario.  In the language of the
$\si$--$\omega$ model this would be expressed through the
momentum--dependence of couplings and wave function renormalizations (form
factors).

Despite the substantial difference in the compression modulus the three
scenarios (A)--(C) all show a similar value of $\olsinuc\simeq30\MeV$ and
therefore a nucleon mass in nuclear matter around $700\MeV$, as supported
by some experimental evidence~\cite{FJL96-1,Gle97-1}.  Also the critical
value $\mu_{\rm eff,nuc}\simeq250\MeV$ is very similar for these three
models. All three scenarios support the existence of a first-order
transition from the vacuum to nuclear matter. Extending this result to
small $T\not=0$ this results in a first-order transition  from a nucleon
gas at low density to a nuclear liquid at high density, in complete
analogy to the vapor-water transition.

\section{Droplet model for nuclei \label{dropmod}}

For a first-order gas-liquid transition we can describe
sufficiently large nuclei
as droplets of nuclear matter in a surrounding vacuum.  For
quantitative estimates of their properties we have to take into account that
because of the surface tension the pressure inside the droplet is different
from zero.  The nucleus is at equilibrium if the pressure equals the
derivative of the sum of surface and Coulomb energy\footnote{We neglect here
the asymmetry effect from the proton--nucleon mass difference.} with respect
to the volume
\begin{eqnarray}
  \label{AAA119}
  &&p=\frac{\prl E_\Si}{\prl V}+\frac{\prl E_{\rm c}}{\prl
    V}\nonumber\\
  &&E_\Si=(36\pi)^{1/3}\Si V^{2/3},\quad E_{\rm c}=\frac{3\alpha}
  {5}\left(\frac{4\pi}{3}\right)^{1/3}\kappa Z^2V^{-1/3}\; .
\end{eqnarray}
The surface tension $\Si$ can be expressed (thin wall approximation)
in terms of the potential as
\begin{equation}
  \label{AAA120}
  \Si=2\left(\frac{Z_\si}{Z_\pi}\right)^{1/2}
  \int^{\si_0}_{\olsinuc}
  d\si\sqrt{2U(\si;\mu)}\; .
\end{equation}
We use the phenomenological relation $Z/B\simeq(2+0.0153B^{2/3})^{-1}$,
$\al=1/137$, and $\kappa$ should be very close to one.  This leads to a
volume-- and therefore baryon number dependent pressure
\begin{equation}
  \label{AAA122}
  p=\frac{4}{3}\left(\frac{3}{\pi}\right)^{1/3}
  q_{\rm nuc}\Si B^{-1/3}-\frac{4}{45\pi^2}
  \left(\frac{3}{\pi}\right)^{1/3}
  \alpha\kappa q^4_{\rm nuc} Z^2 B^{-4/3}
\end{equation}
and a total binding energy per nucleon
\begin{equation}
  \label{AAA124}
  \frac{E_{\rm B}}{B}=
  3\mu(p)-m_{\rm N}-\frac{p}{n}+
  \frac{E_\Si}{B}+\frac{E_{\rm c}}{B}=
  \bt+\frac{E_\Si}{B}+
  \frac{E_{\rm c}}{B}-
  \frac{9}{2K_0}\frac{p^2}{n_{\rm nuc}^2}+
  \ldots\; .
\end{equation}
Neglecting the pressure term the mass formula for nuclei yields the
``experimental'' values
\begin{eqnarray}
  \label{AAA125}
  \ds{\Si^{\rm(n)}} &=& \ds{
    4.22\cdot10^4\MeV^3
    }\nnn
  \ds{\kappa^{\rm(n)}} &=& \ds{
    0.96    }\; .
\end{eqnarray}
For $B=208(12)$ one finds $p=0.9(5.9)\cdot10^6\MeV^4$.  The pressure therefore
contributes to $E/B$ only very little, $\Dt(E/B)=-0.012(-0.53)\MeV$ and can
indeed be neglected for large $B$. For small $B$ eqs.~(\ref{AAA122}),
(\ref{AAA124}) result in an interesting correction to the mass formula,
which is usually not taken into account in the droplet model for nuclei.
Comparison with fig.~\ref{n_p} shows that our model yields a
baryon density which is indeed almost
independent of $B$ for large nuclei. (Note that $\ol{n}^{\rm(n)}$ formally
corresponds to $B\ra\infty$.)
On the other hand, for small nuclei the baryon density is enhanced.
The value of the surface tension as computed from eq.~(\ref{AAA120}) for
different parameter sets can be found in table~\ref{tab:02}.  This result is
quite reasonable in view of the uncertainties, first from the proper choice of
a coarse grained potential in~(\ref{AAA120})
(cf.~\cite{BW97-1,BTW96-1,STW98-1}), and
second from the choice of parameters in $U_0$.  In fact, the successful
explanation of the small ratio $(\Si^{\rm(n)})^{1/3}/q_{\rm nuc}$ is
encouraging.  In summary, our simple approach gives a quite reasonable picture
for nuclei. We emphasize that once $U_0(\sigma)$ is fixed, our approximations
allow for a ``first principle calculation'' of properties of nuclei!

\section{Two flavor quark hadron phase transition}
\label{sec:08}

At the critical chemical potential $\mu_{\rm nuc}$ the free energy
$U(\sigma;\mu)$ shows two degenerate minima: one at $\sigma_0$, with vanishing
density and the other at $\sigma_{\rm nuc}$, where nuclear matter density
$n_{\rm nuc}$ is reached. In this section we consider densities higher than
$n_{\rm nuc}$. For sufficiently high density one may expect, and we observe,
a further transition from nuclear matter to quark matter.
A new first-order phase transition would be
related to a third distinct minimum of the effective potential $U$.

The results of a quantitative analysis for the polynomial
potential~(\ref{AAA030}) with the parameter sets $A$--$C$ are reported in
table~\ref{tab:03}.
\begin{table}[htbp]
  \begin{center}
    \leavevmode
    \begin{tabular}{|c||c|c|c|c|c|c|}
      \hline
      &
      $\ds{\frac{\mu_{\rm qm}}{\MeV}}$ &
      $\ds{\frac{\mu_{\rm eff,qm}}{\MeV}}$ &
      $\ds{\frac{n_{\rm qm}}{n_{\rm nuc_{  }}}}$ &
      $\ds{\frac{\olsi_{\rm qm}}{\MeV}}$ &
      $\ds{\frac{M_{\rm q,qm}^{  }}{\MeV}}$ &
      $\ds{\frac{\Si_{\rm qm}}{10^6\MeV}}$ \\
      \hline\hline
      $A$ &
      $975.5$ &
      $530$ &
      $8.6$ &
      $2.2$ &
      $15.8$ &
      $5.9$
      \\\hline
      $B$ &
      $859.7$ &
      $484$ &
      $6.5$ &
      $2.2$ &
      $15.4$ &
      $4.8$
      \\\hline
      $C$ &
      $511$ &
      $370$ &
      $2.9$ &
      $5$ &
      $34$ &
      $1.7$
      \\\hline
    \end{tabular}
    \caption{\footnotesize Critical quantities for the quark--hadron phase
      transition. The values for the chemical potentials $\mu_{\rm qm}$ and
      $\mu_{\rm eff,qm}$, the baryon density in the quark matter phase $n_{\rm
        qm}$, the order parameter $\olsi_{\rm qm}$, the effective quark mass
      $M_{\rm q,qm}$and the surface tension $\Si_{\rm qm}$ for the sets $A$
      and $B$ should be interpreted as an illustration of the uncertainties of
      a polynomial extrapolation of the potential $U_0$ to the origin
      $\si=0$.}
    \label{tab:03}
  \end{center}
\end{table}
Using for the transition momentum the ansatz of eq. (\ref{eq:jjj001}) 
with $\sigma_c$ and $\sigma_H$
in a reasonable range, we find a first order transition between nuclear and  
quark matter.
The sets $A$ and $B$ with high values of $\mu_0$ and $\ol{m}$ lead,
however, to relatively large values of the critical chemical potential
$\mu_{\rm qm}$ at which the transition from nuclear to quark matter 
occurs. One also finds large values of $\mu_{\rm eff,qm}$ and the 
associated critical
baryon density $n_{\rm qm}$ in the quark matter phase.
This sheds doubts on the reliability of this computation.
One may argue that for such high values of $\mu_{\rm eff}$
there is no good reason why a separate minimum for nuclear
matter should persist.
The prediction of a first order transition for the saturation scenario
(C) seems more robust in this respect.

We next present a short description of the dominant effects
that lead to our picture of a first order quark-hadron phase transition.
This should also give an impression of the substantial uncertainties
still inherent in this picture.
The dominant mechanism for a possible first-order quark-hadron phase
transition is the rapid decrease of $U_\mu$ at $\sigma=0$ due to the
fluctuations of massless  quarks, whereas at the potential minimum
which corresponds to nuclear matter the effect of the quark fluctuations
is reduced by their effective mass and by $q_H>0$.
For $\mu$ increasing beyond $\mu_{\rm nuc}$ the density of nuclear matter
increases beyond $n_{\rm nuc}$ and $\ol{\si}^{\rm(nuc)}(\mu)$ decreases
(scenarios $A$, $B$).  In our crude picture this continues until $\mu_{\rm
  eff}$ reaches the value $\mu_{\rm nf}(\ol{\si}^{\rm(nuc)})$
(cf.~eq.~(\ref{AAA106})). At the corresponding density the equation of
state becomes very stiff, similar to the saturation scenario $C$ discussed in
section~\ref{sec:06} (eq.~(\ref{AAA117})). The density can further increase
because of quark contributions only once $\mu_{\rm eff}$ becomes larger
than $\mu_{\rm q}(\olsi^{\rm(nuc)})$.  (For the parameter set $C$
corresponding to the saturation scenario $\mu_{\rm eff}$ must first
exceed $\mu_{\rm q}(\olsinuc)=334\MeV$ ($\mu>383\MeV$) before the density
can increase beyond nuclear density.)
As a result of the ``frozen density'' the rate of decrease
of $U_\mu$ is also frozen according to eq. (\ref{AAA003}).
On the other hand, the quarks always fully
contribute to $\prl U/\prl\mu$ at $\si=0$
(in the absence of current quark masses). For $\mu_{\rm eff}>q_{\rm H}$
and $\mu_{\rm eff}>\mu_{\rm nf}(\olsinuc)$ the potential at $\si=0$
decreases therefore faster with $\mu$ than for the nuclear matter phase at
$\olsi^{\rm(nuc)}$ (cf.~eqs.(\ref{AAA200}), (\ref{AAA010}) with $q_{\rm
  H}(\si=0)=0$ and $q_{\rm H}(\olsinuc)=q_{\rm H}$).

In the vicinity of $\sigma=0$ the effect of the current quark
masses should be included for a quantitative calculation. They  push
the minimum to positive $\sigma$ such that large enough quark masses
typically destroy a possible first-order transition. In fact, at
sufficiently high $\mu$ the effective potential~(\ref{AAA012}) always has
a new minimum near the origin at
\begin{equation}
  \label{AAA032}
  \ds{\ol{\si}^{\rm(qm)}(\mu)} \simeq \ds{
    \frac{m_\pi^2}{m_0^2(\mu)}\si_0
    }\; .
\end{equation}
Here the mass parameter
\begin{equation}
  \label{AAA101}
  m_0^2(\mu) = \left.\frac{1}{4}
  \frac{\prl^2 U}{\prl\si^2}\right|_{\si=0}=
  \frac{3}{4\pi^2}h_q^2\mu_{\rm eff}^2-\ol{m}^2 \, ,
\end{equation}
\begin{equation}
  \label{AAA102}
  \ol{m}^2\equiv2\si_0^2\left(\la-\gm_3+\gm_4-\gm_5\right)-
  m_\pi^2=
  \frac{\mu_0^4}{2\pi^2\si_0^2}+
  \left(\gm_4-\frac{2}{3}\gm_3-\frac{6}{5}\gm_5\right)
  \si_0^2-3m_\pi^2
  \; .
\end{equation}
corresponds to the curvature of the potential at the origin
and $m_\pi^2 \sigma_0$ reflects the linear source term. At this
minimum the effective quark mass  $m^{\rm(qm)}=h\ol{\si}^{\rm(qm)}=h\si_0
m^2_{\pi}/m_0^2(\mu)$,  vanishes in the chiral limit $m_\pi\to0$
and for $\mu \to \infty$. For small enough $m^{\rm(qm)}$ we
identify the corresponding phase with quark matter.  For a vanishing
current quark mass ($\jmath=0$, $m_\pi=0$) chiral symmetry is restored in
this phase. In case of a first order phase transition and, in particular,
for small current quark masses, one typically finds a situation where two
different local minima at $\sigma^{\rm (qm)}$ and $\sigma^{\rm (nuc)}$ 
coexist.
As $\mu$ increases, the height of the potential for the quark
matter phase $U^{\rm(qm)}(\mu)=U(\ol{\si}^{\rm(qm)}(\mu);\mu)$ decreases
faster than the one for the nuclear matter phase
$U^{\rm(nuc)}(\mu)=U(\olsi^{\rm(nuc)}(\mu);\mu)$, where we remind that
decreasing $U$ corresponds to increasing the pressure $p=-U$.  This can be
seen directly from eqs.~(\ref{AAA011}), (\ref{AAA014a}), since
$\olsi^{\rm(qm)}<\ol{\si}^{\rm(nuc)}$ and $\frac{\prl U}{\prl
  \si}(\olsi;\mu)=0$. One concludes that for large enough $\mu$ the
absolute minimum of $U$ is always given by
eqs.~(\ref{AAA032})--(\ref{AAA102}).

Away from the chiral limit the quark--hadron phase transition is not
characterized by a change of symmetry in our model\footnote{As mentioned in
  the introduction, we do not take into account in the present approach the
  possible spontaneous breaking of color at high density
  \cite{CSC} or in the vacuum \cite{W3F},\cite{BW2F}.}.
It could therefore be of
first order or a crossover. (A second order transition would require an
additional tuning of parameters.)
Our numerical evaluation of $U$ for the three scenarios
A, B, C shows that the existence of a first-order transition
depends on assumptions about $q_H(\sigma)$. For many
reasonable functional forms we find indeed a first-order
transition. In view of the remaining uncertainties
it seems useful to establish general criteria
for the occurrence of a first-order transition within our
computation. Indeed, a first-order transition is guaranteed if
a range of $\mu$ exists for which $m_0^2(\mu)$ is positive and
substantially larger than $m_\pi^2$ whereas $\ol{\si}^{\rm(nuc)}(\mu)$
remains of the same order of magnitude as $\si_0$. In this case one has
$\ol{\si}^{\rm(qm)}(\mu)\ll\ol{\si}^{\rm(nuc)}(\mu)$ and the mass term at
$\ol{\si}^{\rm(qm)}(\mu)$ is well approximated by $m_0^2(\mu)$ and
therefore positive. By definition the mass term at
$\ol{\si}^{\rm(nuc)}(\mu)$ is also positive. Two local minima of $U$
coexist for this range of $\mu$. As $\mu$ is increased further the mass
term at $\ol{\si}^{\rm(qm)}(\mu)$ monotonically grows
(cf.~eq.~(\ref{AAA101})) thus excluding a crossover. Typical values for
$\ol{m}$ from an extrapolation of the polynomial potential~(\ref{AAA030})
for the parameter sets $(A,B,C)$ are $(833.4,709.7,584.9)\MeV$. For these
values a first order transition would be guaranteed for $m_0(\mu_{\rm
  eff})\gta400\MeV$ or $h_q\frac{\mu_{\rm eff}}{260\MeV}>(13.10,11.82,9.92)$
if $\olsi^{\rm(nuc)}(\mu_{\rm eff})$ remains of order $\si_0$.  We use here
a vacuum constituent quark mass of $330\MeV$ or a Yukawa coupling $h_q=7.1$.

In summary, we infer a first-order transition if $\ol{\sigma}^{(\rm nuc)}
(\mu_{\rm eff})$ remains of order $\sigma_0$ for $\mu_{\rm eff}=(480,433,363)$
MeV for the models A, B, C.
For low enough values of $\ol{m}$ (as, for instance, in scenario $C$)
the value of $\mu_{\rm eff}$ is low enough such that
$\ol{\sigma}^{\rm (nuc)}(\mu_{\rm eff})$ is not expected
to be much smaller than the value in nuclear matter
at low pressure, as given in table 2. A first-order
transition occurs then independently of other details of the potential.
On the other hand, for large values of $\mu_{\rm eff}$
the dependence of $\ol{\sigma}^{\rm (nuc)}(\mu_{\rm eff})$ is much
more difficult to assess. It depends crucially
on the way how the quarks are ``switched on'', as expressed in the present
formalism by the functional form of $q_H(\sigma)$. A crossover
or even a rather smooth transition become possible as well.

Actually, a very  natural scenario seems to be a first
order transition at a
critical value $\mu_{\rm eff,qm}$ which is lower than $\mu_{\rm
 q}(\ol{\si}^{\rm(nuc)})$. In this case the quarks do not contribute in
the nuclear matter phase and nucleons are absent in the quark matter phase.
In the following we concentrate on this scenario which can be
realized for the ansatz (\ref{eq:jjj001}) with reasonable values
of $\sigma_c$ and $\sigma_H$.
Typical values of $\mu_{\rm eff,qm}$ for this situation are somewhat above
$q_{\rm H}$, say, $\mu_{\rm eff,qm}\simeq(300-400)\MeV$. The baryon density
in the quark phase at this transition would be around three times nuclear
density. These values occur naturally for values of $\mu_0$ somewhat below
$\mu_{\rm eff,qm}$. An investigation of the coarse grained effective
potential in the framework of the average action for a nonvanishing baryon
chemical potential~\cite{BJW98-2} finds values of $\mu_0$ only slightly
above the constituent quark mass. This can be interpreted as an
information about the potential $U_0$ near the origin and supports
the above scenario.

In order to estimate the critical value $\mu_{\rm eff,qm}$ for the quark
hadron phase transition in this scenario we equate the pressure in the
quark matter phase (for the approximation $m_\pi=0$)
\begin{equation}
  \label{eq:new01}
  p^{\rm(qm)}=\frac{1}{2\pi^2}
  \left(\mu_{\rm eff}^4-\mu_0^4\right)
\end{equation}
with the one in the nuclear matter phase
\begin{equation}
  \label{eq:new02}
  p^{\rm(nuc)}=\frac{2}{\pi^2}q_{\rm H}^3\mu_{\rm eff}-
  \frac{2}{3\pi^2}q_{\rm H}^3
  \sqrt{\ol{m}_{\rm N}^2+q_{\rm H}^2}+
  \ol{p}\; .
\end{equation}
Here $\ol{m}_{\rm N}$ and $\ol{p}$ are the nucleon mass and pressure
corresponding to $\mu_{\rm eff}=\mu_{\rm nf}$, respectively.  For $q_{\rm
  H}$ not much larger than $q_{\rm nuc}$ one may neglect $\ol{p}$.
Inserting two typical sets of values, $\mu_0=320\MeV$, $q_{\rm
  H}=1.05(1.2)q_{\rm nuc}$, $\ol{m}_{\rm N}=0.7(0.6)m_{\rm N}$, one obtains
$\ol{\mu}_{\rm eff,qm}=390(440)\MeV$. This corresponds to a critical baryon
density in the quark matter phase
\begin{equation}
  \label{eq:new03}
  n_{\rm qm}=3.4(4.9)n_{\rm nuc}\; .
\end{equation}

For the scenarios (A) and (B) the existence of two minima found
in our computation may well be an artefact of our inaccurate
treatment of the transition from quarks to nucleons. Indeed, there
are reasonable forms of $q_H(\sigma)$ for which the nuclear matter
minimum has reached small values of $\ol{\sigma}_{\rm nuc}$ already
for substantially smaller $\mu_{\rm eff}$. This would favor
a smooth transition. One should
remember, though, that the estimate of $\mu_{\rm eff}$ depends
crucially on the value of $\ol{m}$ in eq. (\ref{AAA101}),
which is only poorly known. In fact, the
observed meson masses and decays contain only
very limited information about the behavior of $U_0$ near the origin. We do
not expect a polynomial expansion of $U_0$ around $\si_0$ to lead to a very
good approximation of the potential in the vicinity of the origin. It is
certainly possible to extrapolate a form of $U_0$ which is compatible with
nuclear physics constraints in the region $0.6\si_0<\si<1.5\si_0$ to a wide
range of parameters $\mu_0$ and $\ol{m}$ characterizing the behavior of
$U_0$ near $\si=0$. Furthermore, a possible $\si$--dependence of
$g^{(\om)}$ or $M_\om$ would substantially affect the ratio $\mu_{\rm
  eff}/\mu$. In particular, a smaller value of $g^{(\om)}/M_\om$ for the
quarks (near $\si=0$) would enhance the effective chemical potential for
given $\mu$ in the quark phase, thereby shifting the transition to lower
values of $\mu$.  We conclude that the spread in the values in 
table~\ref{tab:03}
(especially those corresponding to sets $A$ and $B$) should be considered
as an illustration of the uncertainties still inherent in the polynomial
extrapolation rather than as actual predictions (which we expect closer to
eq.~(\ref{eq:new03}) in case of a first order transition). 
This uncertainty is reduced significantly once
independent information about the behavior of $U_0$ near $\si=0$ becomes
available as, for example, from ref.~\cite{BJW98-2}.

We have also computed the surface tension $\Si_{\rm qm}$ for the
quark--hadron transition at the critical $\mu_{\rm qm}$.  It turns out to
be much larger than the one between the nucleon gas and nuclear matter.
The quantitative value is given in table~\ref{tab:03}. The surface tension
depends, however, strongly on the details of the transition from quark to
nucleon degrees of freedom (e.g., $\si_{\rm c}$ and $\si_{\rm H}$).
Stability of nuclear matter requires the critical chemical potential
$\mu_{\rm qm}$ for a possible quark--hadron transition to be above
$\mu_{\rm nuc}$ as realized for our parameters.  At the quark hadron phase
transition the quark mass $M_{\rm q,qm}$ in quark matter is much smaller
than in nuclear matter.  Nevertheless, it is substantially larger than the
current quark mass. We quote the value of the order parameter $\olsi_{\rm
  qm}\equiv\ol{\si}^{\rm(qm)}(\mu_{\rm qm})$ for the quark phase in
table~\ref{tab:03} together with the corresponding quark mass. For
$\mu=\mu_{\rm qm}$ the values in the nuclear matter phase are
$\olsi^{\rm(nuc)}(\mu_{\rm qm})\simeq24\MeV$, $M_{\rm
  q,qm}^{\rm(nuc)}\simeq(165-170)\MeV$.  Since $\mu_{\rm qm}$ may exceed
the effective strange quark mass, the strange quarks could play a role for
this transition in real QCD.

To summarize this section, our first computation exhibits
a first-order phase transition between nuclear and quark matter
at a critical density which is a few times nuclear density. 
If such a transition really occurs, our values for the density
seem much more realistic than the very low values typically
obtained in simple quark model computations. This underlines
the importance of the correct treatment of the baryons in the 
nuclear matter phase. We have
also seen, however, that the uncertainties remain very substantial
and a smooth behavior remains also conceivable. The unknowns we have
encountered will be present in any realistic mean field-type
treatment of the transition between quark and nuclear matter. Whereas
high density quark matter can perhaps be dealt with rather reliably
at high enough baryon density, the main problem concerns the
behavior of nuclear matter at high density for which confinement
effects (the binding to baryons) cannot be neglected. Any discussion
of a phase transition needs knowledge about both phases concerned.
The high density nuclear matter phase therefore needs always to
be understood quantitatively and a pure quark model cannot give a reliable
description (unless the ``binding of baryons'' is somehow incorporated).
This raises substantial doubts about the applicability of many
mean field statements about this transition in simple quark models.

\section{Conclusions}
\label{Concl}

We have presented here a new method for the computation of the dependence
of the free energy on the chemical potential.  It is based on an
approximate solution to an exact functional differential equation. This
method allows us to put mean field theory into a more systematic context.
Chiral symmetry is explicitly implemented and phenomenological information
about pion masses and decay constants is taken into account. Expressions
which are close to mean field theory describe the difference in the free
energy between vanishing and non--vanishing chemical potential. They can be
considered as the leading order in a series of systematic truncations of
the exact differential equation~(\ref{AAA004}). On the other hand, the free
energy for vanishing chemical potential is not reliably described by mean
field theory. Many relevant characteristics of this quantity can, however,
be inferred from observation.

Perhaps the most important new feature in our approach is that
quark and nucleon fluctuations can be treated simultaneously
within the same computation of the free energy. Only this allows
a simultaneous description of the nuclear gas-liquid transition
and the transition from nuclear to quark matter. A method which
can deal both with nucleons and quarks is crucial for any quantitative
treatment of a possible phase transition from nuclear to
quark matter. Actually, for this transition 
the main difficulty lies in the
understanding of the ``low density phase''. This phase is nuclear
matter at a critical density of perhaps several times nuclear
density, where standard nuclear physics is not of much help and
the binding of quarks into nucleons nevertheless remains an important
ingredient. Simple quark models not accounting for this binding,
like NJL-type models, are insufficient for a description of this
transition.

We take here a very simple approximation where we
use quark and nucleon degrees of freedom in their appropriate momentum
ranges. For high momenta, $\vec{q}^{\,2}>q_{\rm H}^2$, the quark meson
model gives a useful approximation. For small momenta,
$\vec{q}^{\,2}<q_{\rm H}^2$, the effects of confinement have to be taken
into account and we describe the carriers of baryon number as nucleons. Our
simple model leads to a unified description of the nucleon gas, nuclear
matter and transition to quark matter.
The appearance of three phases of strongly
interacting matter is related to three distinct minima of the effective
potential for the $\si$--field. (Typically only two coexist
simultaneously.) This rich structure is a consequence of the fact that more
energy levels fall below the Fermi energy for nucleons than for quarks.
This results in a substantial enhancement of the density 
or, equivalently, the $\mu$-dependence of the
free energy due to low momentum nucleon fluctuations.
The enhancement of the effect of a Fermi-gas of massive nucleons as
compared to a gas of quarks with constituent masses only about
a third of the nucleon mass is actually a huge factor of 27.
This large enhancement relies only on the different effective
masses and, therefore, Fermi surfaces of nucleons and quarks.
Consequently, this property
is rather independent of other more detailed features of the model.

The use of nucleon degrees of freedom and the corresponding large
density enhancement factor as compared to quark degrees of freedom
is crucial for a any realistic description of the liquid--gas
nuclear transition.  It also shifts the transition from nuclear to quark
matter to a larger chemical potential and baryon density, as
compared to a description of the fermionic fluctuations in terms of quarks
alone. This results in a substantial pressure at the coexistence between quark
and nuclear matter. Standard
Nambu--Jona-Lasinio type quark models fail to take into
account this ``density enhancement'' and, therefore, typically 
predict a transition to quark matter at too low densities.

The gas--liquid nuclear transition can be described within the present
approach in a quantitative way, at least for low temperature.
We present here two classes of models, one similar
to the $\sigma-\omega$-model widely used in nuclear physics,
the other a ``saturation model'' where the short distance repulsion
between nucleons arises not only from vector-meson exchange but
also reflects the transition to quark degrees of freedom at
high density. For a suitable choice of parameters of the effective
meson masses and interactions in vacuum our models give a
successful description of the nuclear droplet model, consistent with the
observed nuclear density, the binding energy per nucleon, the compression
modulus and the nucleon mass in nuclear matter. They explain why nuclear
density is approximately independent of the baryon number of a nucleus. We
also obtain a realistic value for the nuclear surface tension and we have
computed small corrections to the baryon density and the mass formula for
nuclei due to nonvanishing pressure. Isospin violation and electromagnetism
can be incorporated easily in our model. This should give a reliable
equation of state for neutron stars in the region of moderate densities.
Our model predicts coupling constants which directly enter the effective
nucleon--nucleon potential. Comparison with nucleon scattering experiments
will provide an interesting test in the future.

A potential shortcoming of these models are the large
meson self-interactions in vacuum which are needed for a
realistic description. Within a linear $\sigma$-model they
correspond to large coefficients of terms $\sim \sigma^6,
\sigma^8$ or $\sigma^{10}$ which have not been found in
previous renormalization group studies. It remains to be seen if
a more complex structure of vacuum expectation values in the
nuclear matter phase and in the vacuum -- including spontaneous
breaking of color -- can ease this problem. The generalization
of our approach to additional order parameters is straightforward.
One may consider our quantitative results as a ``prototype
calculation'' for understanding the properties of nuclei
from QCD. It can be adapted to more complex settings. Particularly
important will be a reliable computation of the vacuum-effective
potential.

Within our approximations we
find a first order phase transition between nuclear and quark matter at
high density and vanishing temperature. The critical density is
typically three to five times nuclear matter density.
We emphasize, however, that some
important information is still missing for a quantitative understanding of
the quark--hadron transition: The first problem concerns the appropriate
formulation of a coarse grained effective potential and a determination of
its shape for the vacuum. This is needed since first-order transitions
require a non-convex coarse-grained potential whereas the
inclusion of long wavelength fluctuations leads to a convex
effective potential.
Within the linear quark meson model we have addressed this issue
in the context of the average action~\cite{BJW98-2}.  The second loose end
is a more detailed understanding of the change from quark to baryon
effective degrees of freedom. This concerns primarily the behavior of
nuclear matter at densities much larger than nuclear density.  It is
therefore very relevant for a quantitative description of the quark--hadron
transition. Furthermore, the neglected strange quarks could play
a relevant role at sufficiently high density.
Finally, the presence of additional ``color symmetry breaking''
expectation values are a crucial ingredient for the understanding
of states with high baryon density. Again, this could 
influence the nature of the transition from nuclear to quark matter 
\cite{ARRW}.
In view of all these uncertainties we conclude that no definite
statements about the nature of this transition can be made at present.
Both a genuine first-order phase transition and a relatively
smooth change remain possible. The uncertainties mentioned here are
common for other analytical approaches and constitute
a major difficulty for a quantitative understanding of the
quark--hadron phase transition.

\bibliography{habref}

\end{document}